\documentclass[aps,pra,preprint,groupedaddress]{revtex4-2}

\usepackage{color}
\usepackage{placeins}
\usepackage[utf8]{inputenc}
\usepackage{graphicx}
\usepackage{gensymb}
\usepackage{amsmath,amssymb}
\usepackage{placeins}
\usepackage{subcaption}
\usepackage{multirow}
\usepackage[table,xcdraw]{xcolor}
\usepackage{booktabs}

\newcommand{\Christoffel}[3][]{\genfrac{\{}{\}}{0pt}{}{#2}{#3}_{#1}}
\begin{document}


\title{Active Flow Control for Drag Reduction of a Plunging Airfoil under Deep Dynamic Stall}


\author{Brener L. O. Ramos}
\email[]{brener.lelis@gmail.com}
\affiliation{University of Campinas, Campinas, SP, Brazil}

\author{William R. Wolf}
\email[]{wolf@fem.unicamp.br}
\affiliation{University of Campinas, Campinas, SP, Brazil}

\author{Chi-An Yeh}
\email[]{cayeh@seas.ucla.edu}
\affiliation{University of California, Los Angeles, CA, USA}

\author{Kunihiko Taira}
\email[]{ktaira@seas.ucla.edu}
\affiliation{University of California, Los Angeles, CA, USA}



\date{\today}

\begin{abstract}
High-fidelity simulations are performed to study active flow control techniques for alleviating deep dynamic stall of a SD7003 airfoil in plunging motion. The flow Reynolds number is $Re=60{,}000$ and the freestream Mach number is $M=0.1$. Numerical simulations are performed with a finite difference based solver that incorporates high-order compact schemes for differentiation, interpolation and filtering on a staggered grid. A mesh convergence study is conducted and results show good agreement with available data in terms of aerodynamic coefficients. Different spanwise arrangements of actuators are implemented to simulate blowing and suction at the airfoil leading edge. We observe that, for a specific frequency range of actuation, mean drag and drag fluctuations are substantially reduced while mean lift is maintained almost unaffected, especially for a 2D actuator setup. For this frequency range, 2D flow actuation disrupts the formation of the dynamic stall vortex, what leads to drag reduction due to a pressure increase along the airfoil suction side, towards the trailing edge region. At the same time, pressure is reduced on the suction side near the leading edge, increasing lift and further reducing drag. 
\end{abstract}


\maketitle


\section{Introduction}

Unsteady flows over plunging and pitching airfoils with large excursions in effective angle of attack exhibit the phenomenon of dynamic stall. This process is characterized by unsteady separation and formation of a large leading-edge vortex that exerts high amplitude fluctuations in the aerodynamic loads. Comprehensive reviews of this phenomenon in the context of helicopter rotor blades and pitching airfoils are provided by \cite{McCroskey:1982,Carr:1988,Platzer:1988,Corke:2015}.
For the case of flapping wings, as well as for severe impinging gusts, highly unsteady forcing induces the formation of dynamic stall including a leading-edge vortex \cite{Eldredge:2019}. The evolution and interaction of such vortical structures with the aerodynamic surfaces have a significant impact on flight stability and performance.  At certain conditions, dynamic stall can lead to negative damping that results in a limit-cycle growth of rotor displacements. This phenomenon is referred to as stall flutter \cite{Ham:1966} and it can lead to catastrophic mechanical failure. 

Although several studies have been conducted for pitching airfoils at high Reynolds numbers, research on dynamic stall for plunging airfoils is more scarce, especially at low and moderate Reynolds numbers. The study of airfoils in plunging motion finds application in design and operation of small unmanned air vehicles and micro air vehicles and, therefore, we aim to extend our knowledge on the flow features involved in fully separated low Reynolds number flows involving deep dynamic stall.
	
High-fidelity simulations can provide an abundance of data with both high spatial and temporal resolutions. For example, several two-dimensional computational studies are available in the literature regarding dynamic stall under laminar, transitional, and turbulent flow conditions \cite{Platzer:1988,Visbal:1989,Visbal:1991,Visbal:1994,Visbal:1990,Radespiel:2007}. For high Reynolds number flows, numerical simulations traditionally employ a hierarchy of turbulence models augmented in some instances with empirical transition predictions. Recently, Visbal and co-authors have employed implicit large eddy simulation (ILES) to investigate the phenomenon of dynamic stall for different flow configurations including plunging and pitching motion \cite{Visbal:2011,Visbal:2014,Visbal:2015,Visbal:2017,Benton:2018,Visbal:2018}. 
	
In the present work, we perform implicit large eddy simulations to study the flow physics of deep dynamic stall over a plunging SD7003 airfoil. The deep stall regime is characterized by a separation region on the order of the airfoil length while the light stall regime presents a separation region that extends approximately by the airfoil thickness, with a less severe lift loss \cite{McCroskey:1981}. The flow condition investigated is selected based on the availability of results from other high fidelity simulations \cite{Visbal:2011} and particle image velocimetry (PIV) \cite{Kang:2009,Baik:2009,Ol:2009}. A compressible formulation is adopted since the local Mach number near the leading edge of a moving airfoil can be three to five times higher than that in static condition \cite{McCroskey:1981,McCroskey:1982}. As a result, compressibility effects must be taken into consideration even for low Mach number flows. 

Several investigations of dynamic stall control by both active and passive means, especially for pitching airfoils, are described in the survey by \cite{Lorber:2000}. Several control strategies have been tested including leading-edge blowing \cite{Greenblatt:2001,Sun:1999,Weaver:2004}, leading-edge plasma actuation \cite{Corke:2011, Lombardi:2013, Post:2006},  thermo-acoustic actuation \cite{Benton:2018}, vortex generators \cite{Heine:2013, Martin:2008, Traub:2004} and synthetic jets \cite{Ekaterinaris:2002, Florea:2003, Traub:2004}. In some cases, fixed-wing devices have been used, such as slots \cite{Carr:2001}, leading-edge droop \cite{Chandrasekhara:2004, Joo:2006, Martin:2008} and trailing-edge flaps \cite{Feszty:2004, Gerontakos:2006}.

In this work, blowing and suction actuation is modeled at the airfoil leading edge aiming to reduce the overall drag through modification of the dynamic stall vortex. Active flow control strategies by means of periodic forcing can have effects such as attaching otherwise separated flows or avoiding separation, and increasing lift \cite{Greenblatt:2010}. Previous works show that small disturbances can have a considerable impact on the flow dynamics for a pitching NACA0012 experiencing dynamic stall at high Reynolds numbers \cite{Visbal:2014,Visbal:2015,Visbal:2017}. In the current investigation, it is shown that, for a specific frequency range of actuation, drag is substantially reduced while lift is maintained almost unaffected. The physical mechanisms responsible for the changes in the flow field achieved by actuation are then discussed.

\section{Theoretical and Numerical Methodology} 

\subsection{Governing Equations}

To simulate the flow around a moving airfoil, we solve the weakly conservative form of the Navier-Stokes equations in a non-inertial frame. In this form, source terms emerge from grid curvature and frame movement \cite{Warsi:1978, Yamamoto:2001, Orlandi:1989, Choi:1992, Yang:2001}. Here, all terms are solved in contravariant form to allow the use of a curvilinear coordinate system $\{\xi^1,\xi^2, \xi^3\}$. All equations are non-dimensionalized by freestream quantities such as density $\rho_{\infty}$ and freestream speed of sound $c_{\infty}$. Although the Navier-Stokes equations are non-dimensionalized by speed of sound, displayed results and parameters are provided non-dimensionalized with respect to freestream velocity $U_{\infty}$ in accordance with \cite{Visbal:2011}. All length scales are made non-dimensional by the airfoil chord $L$. For a frame of reference with varying velocity in the Cartesian $y$-direction, continuity, momentum and energy equations reduce to
\begin{equation}
\frac{\partial}{\partial t}(\sqrt{g} \rho) + \frac{\partial}{\partial \xi^i} (\sqrt{g} \rho u^i) = 0
\mbox{ ,}
\label{eq:continuity_curv3}
\end{equation}
\begin{eqnarray}
\frac{\partial}{\partial t}(\sqrt{g} \rho u^i) +
\frac{\partial}{\partial \xi^j} \left[ \sqrt{g} \left( \rho u^i u^j - \tau^{ij} + g^{ij} p \right) \right] \nonumber \\
  + \Christoffel{i}{jk}
 \sqrt{g} \Big( \rho u^k u^j + g^{jk} p - \tau^{kj} \Big) =   \sqrt{g} \rho \ddot{h}^i
\label{eq:momentum_curv3}
\mbox{ ,}
\end{eqnarray}
and
\begin{eqnarray}
\frac{\partial}{\partial t} (\sqrt{g} E) +
\frac{\partial }{\partial \xi^j}  \big\{  \sqrt{g} \big[ (E+p) u^j  - \tau^{ij} g_{ik} u^k \nonumber \\ - \frac{ \mu \, M }{Re \, Pr}  g^{ij} \frac{\partial T}{\partial \xi^i} \big] \big\} = \rho \sqrt{g} ( h^j + u^j ) g_{jp}\ddot{h}^p  \mbox{ .}
\label{eq:energy_curv3}
\end{eqnarray}

In order to close the above system of equations the following relations are employed
\begin{equation}
E = \frac{p}{\gamma - 1} + \frac{1}{2} \rho u^i g_{ij} u^j + \frac{1}{2} \rho \dot{h}^i  g_{ij} \dot{h}^i 
\mbox{ ,}
\end{equation}
\begin{equation}
\tau^{ij} = \frac{\mu \, M}{Re} \left( g^{jk} u^i_{\ | k} + g^{ik} u^j_{\ |k} - \frac{2}{3} g^{ij} u^k_{\ |k} \right)
\mbox{ ,}
\end{equation}
and
\begin{equation}
h = h_o \sin(kt) \mbox{ ,} \label{eq:plunge}
\end{equation}
where, $\rho$ represents the density, $u^i$ the $i$-th component of the contravariant velocity vector and $p$ is the pressure. The term $h$ is the frame position (cross-stream motion of the plunging airfoil), $E$ is the total energy, $\mu$ is the dynamic viscosity, $T$ is the temperature, $ k  = \frac{2\pi f U_{\infty}}{L}$ is the reduced frequency, $Re = \frac{\rho U_{\infty} L}{\mu} $ is the chord-based Reynolds number, $M = \frac{U_{\infty}}{c_{\infty}} $ is the freestream Mach number and $Pr$ is the Prandtl number. The dots represent temporal derivatives of the frame position, i.e., frame velocity and acceleration. In the previous equations, covariant and contravariant metric tensors are defined, respectively, as
\begin{equation}
g_{ij} \triangleq \frac{\partial x^{k}}{\partial \xi^{i}} \frac{\partial x^{k}}{\partial \xi^{j}}
\mbox{ ,}
\end{equation}
and
\begin{equation}
g^{ij} \triangleq \frac{\partial \xi^{i}}{\partial x^{k}} \frac{\partial \xi^{j}}{\partial x^{k}}
\mbox{ ,}
\end{equation}
with
\begin{equation}
g = |g_{ij}|  = \left( \frac{\partial x^i}{\partial \xi^j} \right)^2 \mbox{ .}
\end{equation}
The terms $\Christoffel{i}{jk}$ represent the Christoffel symbols of the second kind and details about the present formulation can be found in \cite{Aris:1989}.

\subsection{Numerical Methods}

	A compact sixth-order finite-difference scheme constructed for a staggered grid is used to calculate all spatial derivatives. To determine $f' $ for a given $ f $, a tridiagonal system is solved as
	\begin{align}
	&\alpha f'_{i-1} + f'_i + \alpha f'_{i+1} = b\frac{f_{i+3/2}-f_{i-3/2}}{3\Delta x} \nonumber \\
	& + a \frac{f_{i+1/2}-f_{i-1/2}}{\Delta x} \mbox{ ,}
	\end{align}
	where $\alpha = 9/62$, $a=\frac{3}{8}(3-2\alpha)$ and $b=\frac{1}{8}(-1+22\alpha)$.
	To minimize errors from unresolved scales, a sixth-order compact low-pass filter is applied according to
	\begin{align}
	&\bar{\alpha} \bar{f}_{i-1} + \bar{f}_{i} + \bar{\alpha} \bar{f}_{i+1} = \bar{a} f_i + \frac{\bar{b}}{2} (f_{i+1} + f_{i-1}) \nonumber \\ 
	&+ \frac{\bar{c}}{2}(f_{i+2}+f_{i-2}) + \frac{\bar{d}}{2}(f_{i+3}+f_{i-3}) \mbox{ ,}
	\end{align}
	where, $\bar{a}=\frac{1}{16}(11+10\bar{\alpha})$, $\bar{b}=\frac{1}{32}(15+34\bar{\alpha})$, $\bar{c}=\frac{1}{16}(-3+6\bar{\alpha})$ and $\bar{d}=\frac{1}{32}(1-2\bar{\alpha})$. In the current implicit large eddy simulations, we use $\bar{\alpha}=0.46$, which implies a filter that only acts on poorly resolved high wavenumbers. Therefore, this filter provides a reliable alternative to a SGS model as discussed by \cite{Visbal:2018}.
	Due to the staggered grid, interpolations are necessary to evaluate properties at specific grid locations. To maintain schemes with high-order, a sixth-order interpolation based on finite differences is used according to
	\begin{align}
	& \tilde{\alpha} \tilde{f}_{i-1} + \tilde{f_i} + \tilde{\alpha}\tilde{f}_{i+1} = \frac{\tilde{b}}{2}(f_{i+3/2}+f_{i-3/2}) \nonumber \\
	& + \frac{\tilde{a}}{2} (f_{i+1/2}+f_{i-1/2}) \mbox{ ,}
	\end{align}
	where $\tilde{\alpha}=3/10$, $\tilde{a}=\frac{1}{8}(9+10\tilde{\alpha})$ and $\tilde{b}=\frac{1}{8}(6\tilde{\alpha}-1)$. Additional details on the finite-difference schemes used for derivation, filtering and interpolation can be found in \cite{Lele:1992,Nag:2004}. 
	
	Near the far-field boundaries, a numerical sponge is used to damp acoustic waves. At the inlet and outlet boundaries, a Riemann invariant transformation is implemented as the far-field condition. The airfoil surface is modeled by a no-slip adiabatic wall. Derivatives of inviscid fluxes are obtained by forming fluxes between the grid nodes, on the staggered grid, and differentiating each component. Viscous terms are obtained by first computing the derivatives of primitive variables at their respective locations (see \cite{Nag:2004} for details). Components of the viscous fluxes are then constructed at each node and differentiated by a second application of the compact scheme. Airfoil movement is added through source terms shown in the formulation section. All schemes discussed are implemented with periodic boundary conditions in the spanwise $ \xi^3 $direction. Since we employ an O-grid, periodic conditions are also enforced along the $ \xi^1 $ direction, along the mesh branch cut, where grid points are coincident.
	
	Two time-marching methods are utilized to advance the flow in time. A compact storage explicit third-order Runge-Kutta scheme is used away from solid walls. In the near-wall region, a second-order implicit time marching scheme with approximate factorization derived from the Beam and Warming method is employed. This formulation avoids time step restrictions typical of wall-normal mesh refinement. An overlap layer is applied at the interface between explicit and implicit time marching schemes. The low-pass compact filter is applied after each time-step of both schemes. More details about the numerical framework employed can be found in \cite{Nag:2004}.

\subsection{Actuator Setup}
\label{control}

	In the current work, we perform flow control using blowing and suction  on the leading edge of the airfoil. To simulate an actuator of length $ s \approx 0.01 L$, as shown in Fig. \ref{actuator_location},	a jet velocity is introduced at the actuator location, which is centered around the airfoil leading edge and is imposed with Eqs. \ref{eq:ujet1} -- \ref{eq:ujet3} as
	\begin{align}
	\frac{U_{jet}}{U_{\infty}} & = \frac{U_{jet \hspace{2px} max}}{U_{\infty}} \hspace{2px} F(s) \hspace{2px} G(t) \hspace{2px} P(z) \,\,\,\,\,\,\,\mbox{ with} \label{eq:ujet1} \\
	F(s) 	 & = \exp{ \bigg( -  \frac{(s^{*}-0.01)^2} { 4.5 }  \bigg)},  \hspace{10px} s^{*} = 5 ( s - 0.005 ) \,\,\,\,\,\,\,\mbox{ and} \label{eq:ujet2} \\ 
	G(t) 	 & = \sin(St \hspace{2px} 2 \pi  t) \mbox{ ,}  \label{eq:ujet3}
	\end{align}
where the Strouhal number is $St = \frac{f L}{ U_{\infty} }$.
	\def\temp{0.45}
	\begin{figure}[ht]
		
		\begin{subfigure}[t]{\temp \textwidth}
			\centering
			\includegraphics[width=\textwidth, trim = {0 25px 0 0}, clip]{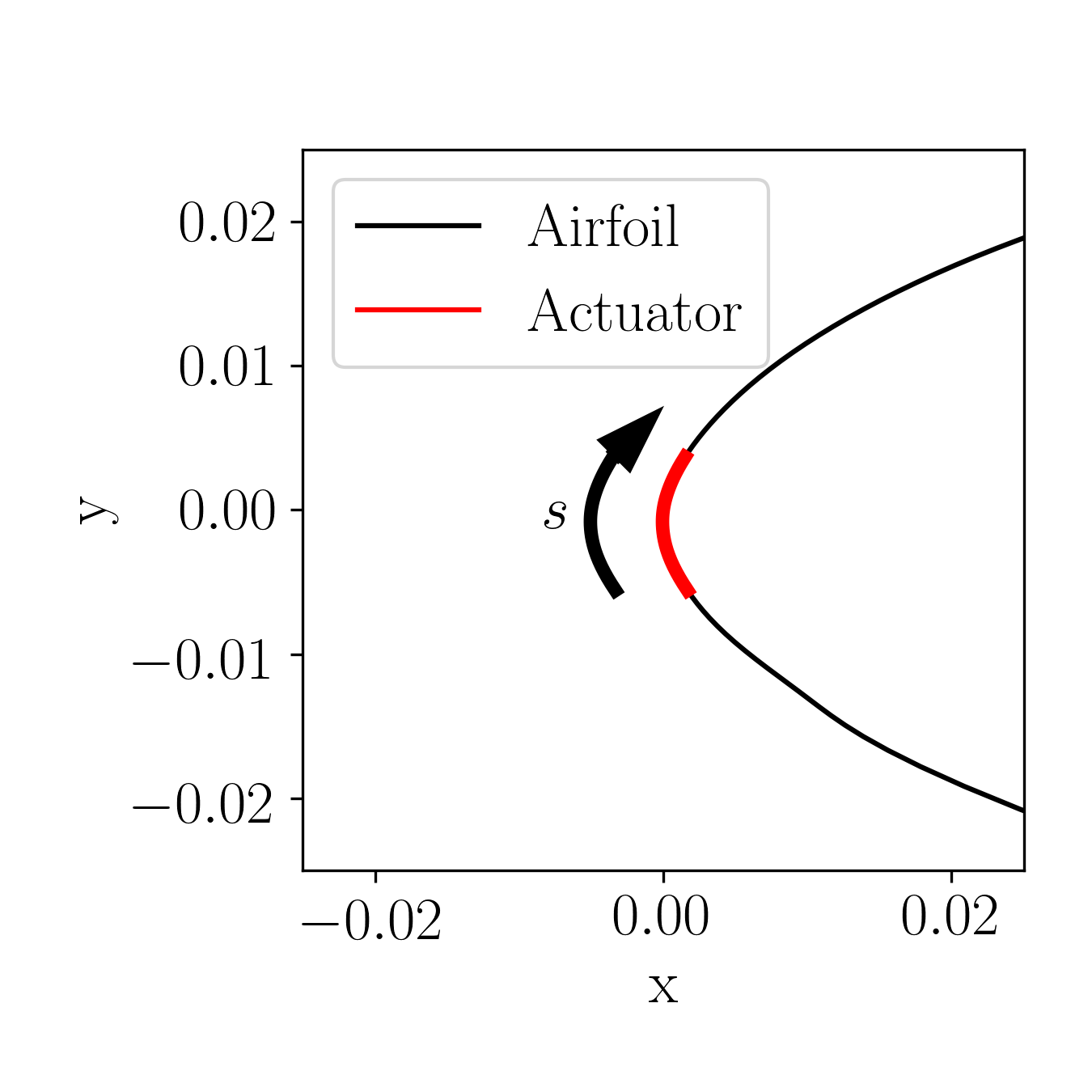}
			\caption{Actuator location in the x-y plane.}
			\label{actuator_location}
		\end{subfigure}		
		\begin{subfigure}[t]{\temp \textwidth}
			\centering
			\includegraphics[width=\textwidth, trim={ 5px 80px 5px 10px}, clip]{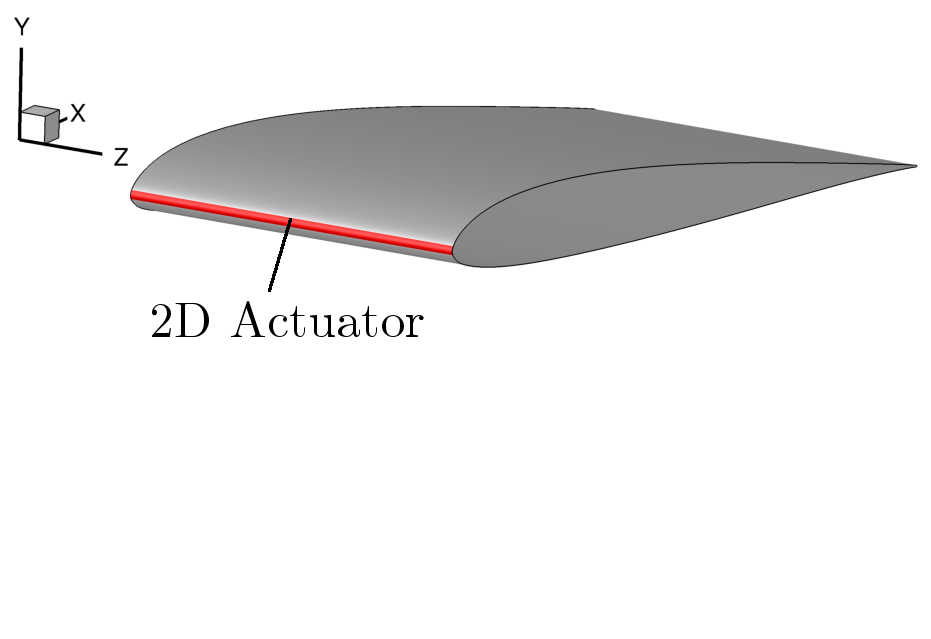}
			\caption{3D view of 2D actuator.}
			\label{actuator_sw_profile}
		\end{subfigure}
		\caption{Actuator setup.}
	\end{figure}
	
	The jet actuation is a sinusoidal temporal function $G(t)$ given by a Gaussian profile $F(s)$ along the wall-tangential direction $ s $ and a profile $ P(z) $ along the airfoil span with maximum jet velocity set as $U_{jet \hspace{2px} max}$. The spanwise actuation functions are chosen with the intent of approximating the format of real slots on the airfoil surface. This would allow comparisons to experiments.
	We defined the actuator chordwise location after analyzing how efficiently the shear layer and overall flow are disturbed with different actuator positions. For a pitching airfoil, Benton et al. showed that an actuator placed near the leading edge effectively modifies the flow with minimum input \cite{Benton:2018}.

\begin{figure}
		\centering
		\begin{subfigure}{0.49\textwidth}
			\centering
			\includegraphics[width=\textwidth,trim={5px 20px 5px 5px}, clip]{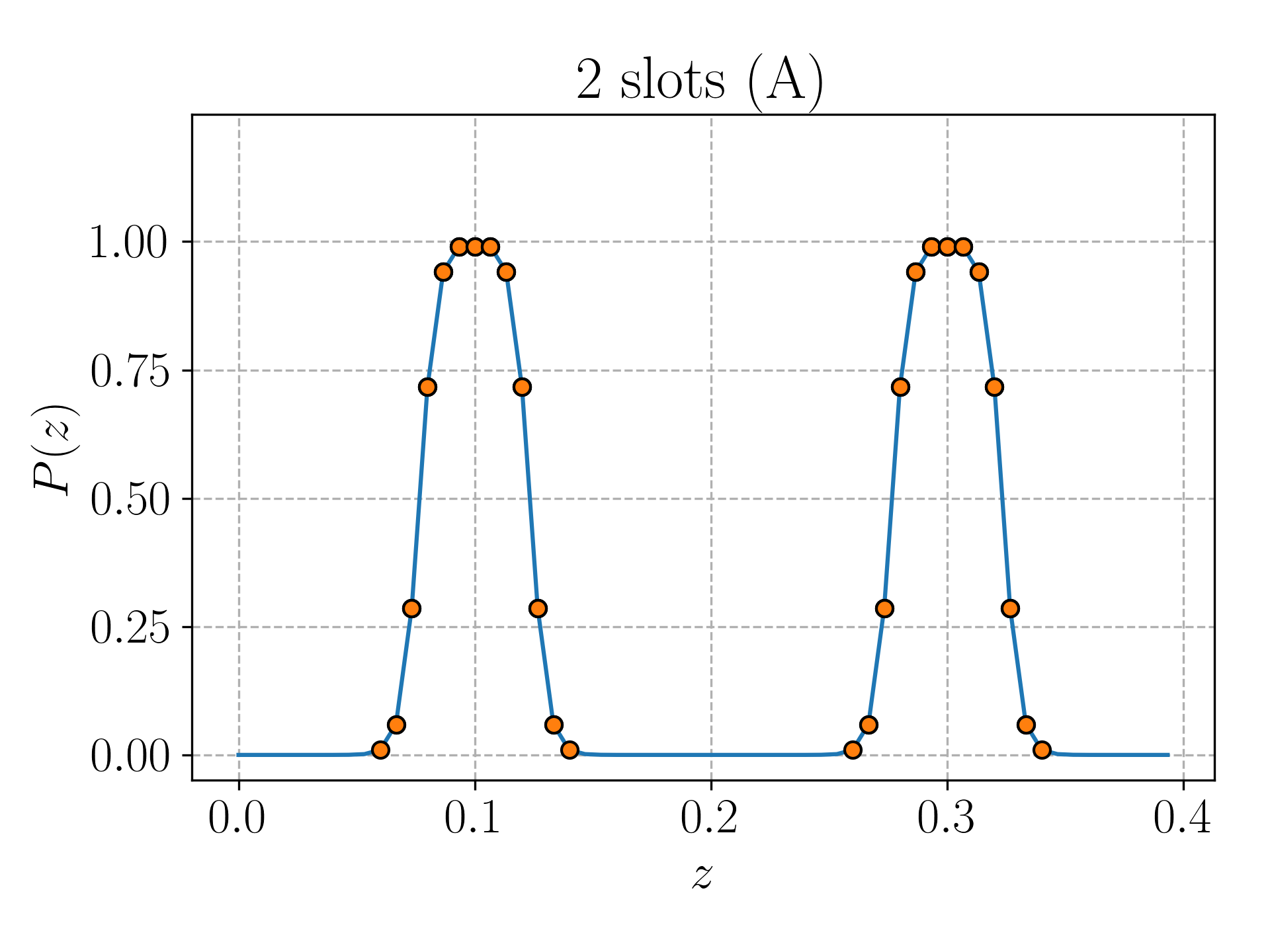}
		\end{subfigure}
		\begin{subfigure}{0.49\textwidth}
			\centering
			\includegraphics[width=\textwidth,trim={5px 20px 5px 5px}, clip]{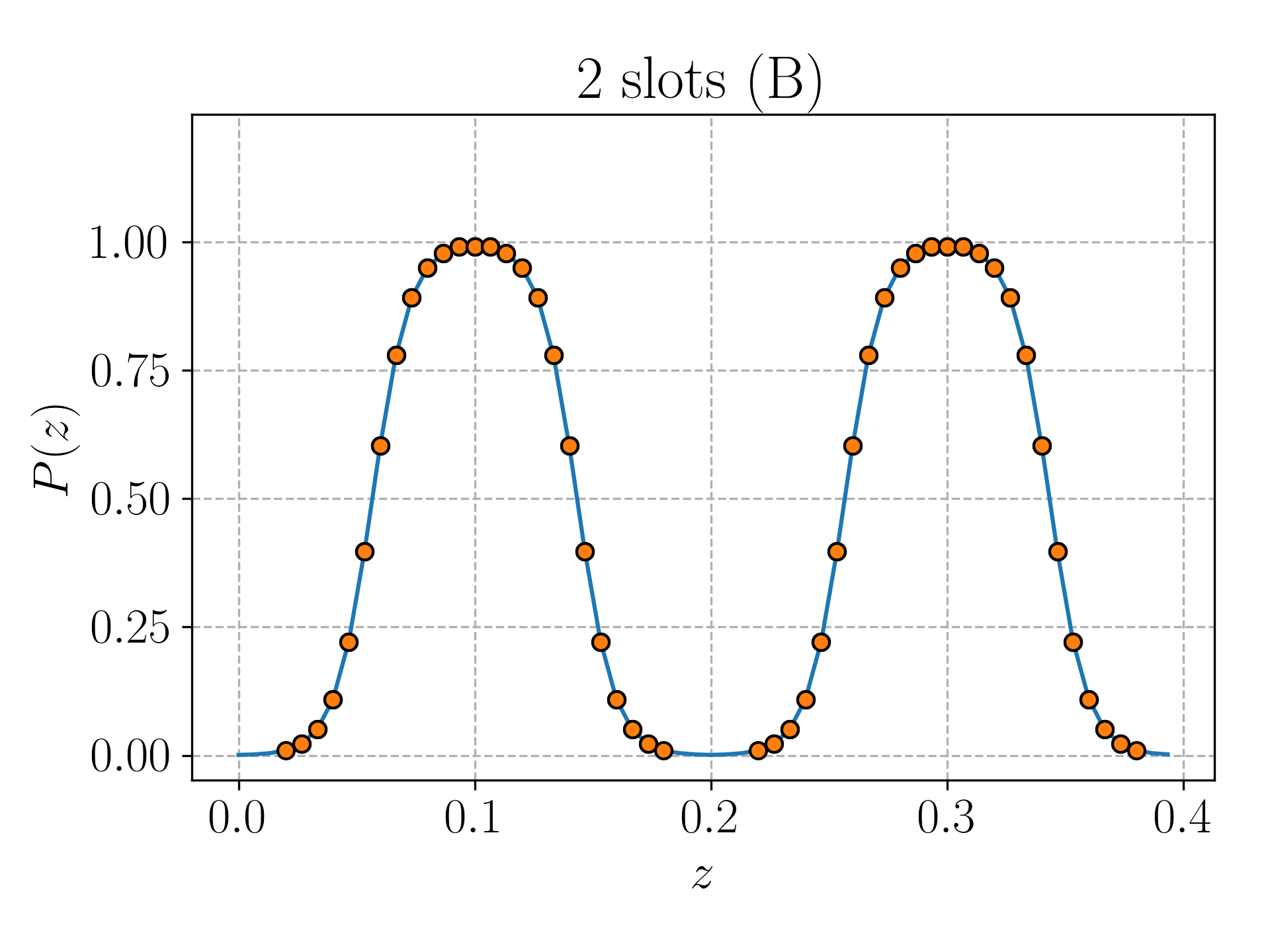}
		\end{subfigure}
		
		\begin{subfigure}{0.49\textwidth}
			\centering
			\includegraphics[width=\textwidth,trim={5px 20px 5px 5px}, clip]{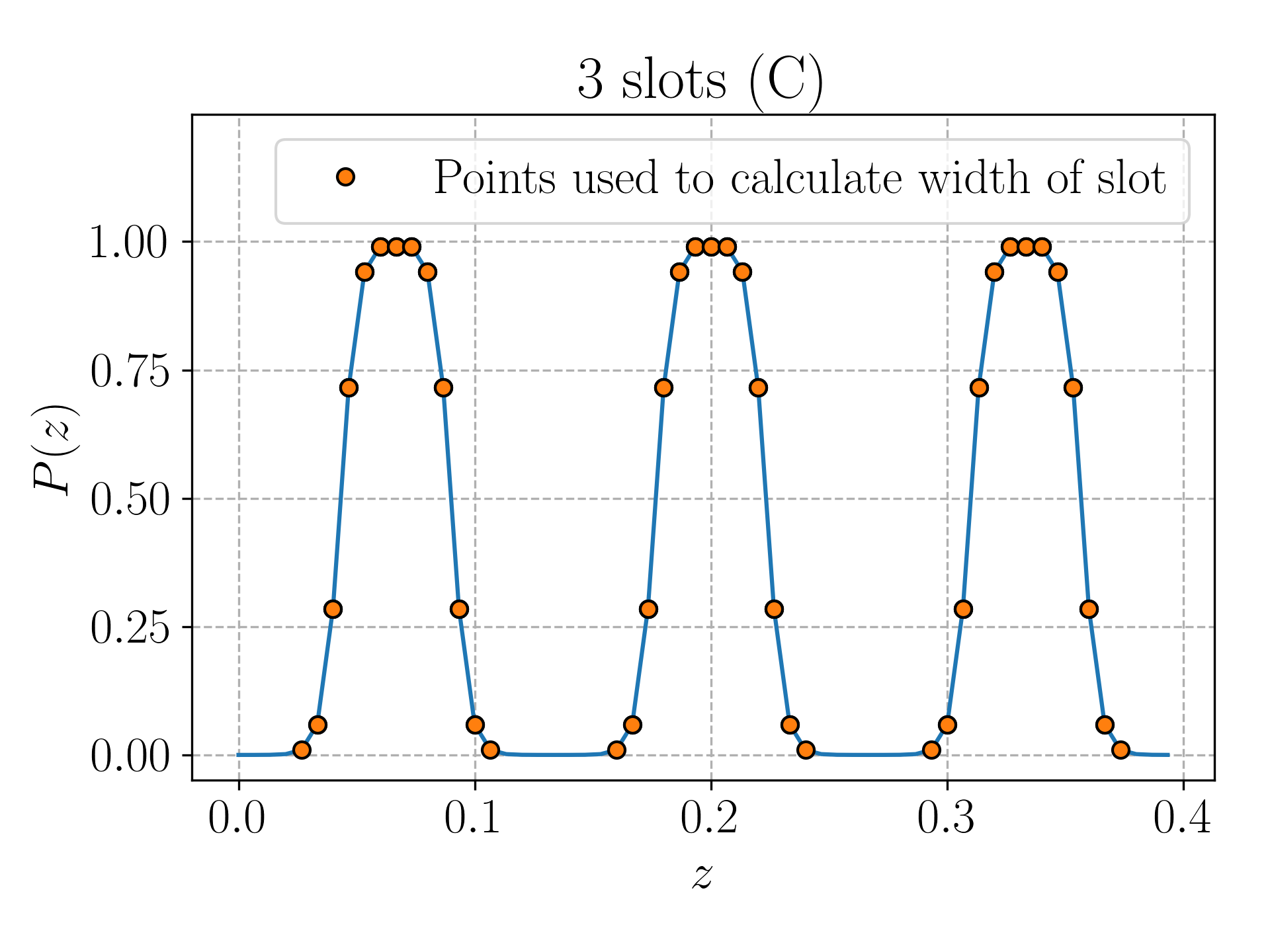}
		\end{subfigure}
		\caption{Profiles of function $ P(z) $ that specifies the spanwise arrangement of actuation.}
		\label{fig:3dprofiles}
	\end{figure}

	To assess the influence of spanwise arrangement of actuation, different spanwise jet configurations are tested through modifications of function $ P(z) $. A 2D actuator is analyzed setting $ P(z) = 1 $ (see Fig. \ref{actuator_sw_profile}). These configurations are obtained appending points according to 
	\begin{align}
		& P_{actuator}(z)  = \tanh \Bigg( \frac{2 (\beta - \alpha)}{\Delta z_{actuator}} z + \alpha \Bigg) \frac{1}{2} + \frac{1}{2},  \hspace{10px} 0 \le z \leq \frac{\Delta z_{actuator}}{2} \label{eq:3dact}
	\end{align}
	with their mirrored values. The profiles are then appended until the whole span is covered as can be seen in Fig. \ref{fig:3dprofiles}. 
	
	In total, four configurations are tested being two consisting of two slots, one with three slots and one two-dimensional actuator. The configurations with two slots have either narrow (A) or wide (B) spanwise jets. The same narrow jets from configuration (A) are tested in the setup with three slots along the airfoil span. Further details about the 3D actuators used in this work are summarized in Table \ref{table:slotsparams}. 
	\renewcommand{\arraystretch}{1.2}	
	\setlength{\tabcolsep}{12pt}	
	\begin{table}[h]
		\caption{Parameters of the 3D actuators from Eq. \ref{eq:3dact}. Coefficients $ \alpha $ and $ \beta  $ are numerical parameters which control the smoothness and stretching of $ P_{actuator}(z) $. }
		\vspace{-13px}
		\begin{tabular}{cccc}
			\toprule \toprule
			& \textbf{2 slots (A)} & \textbf{2 slots (B)} & \textbf{3 slots (C)} \\ \toprule
			\textbf{$\Delta z_{actuator}$} & $ \approx 0.08 L$             & $\approx0.16 L$             & $ \approx 0.08 L$         \\ 
			\textbf{$\alpha$}  & -10.62               & -3.58                & -6.01            \\ 
			\textbf{$\beta$}   & 2.31                 & 2.31                 & 2.31             \\ \bottomrule \bottomrule
		\end{tabular}
		
		\label{table:slotsparams}
	\end{table}
	
	Simulations with actuation frequencies of $ St \in [0.5, 25]$ are first performed for the 2D actuator with the objective of understanding flow response with respect to this parameter. In order to quantify jet actuation efforts, the coefficient of momentum is calculated according to 
	\begin{align}
	C_{\mu} = \frac{  \frac{1}{T_g}\int_{0}^{T_g} \int_{s_0}^{s_n} \int_{0}^{z_{span}} \rho_{\infty} U_{jet}(s,z,t)^2 \hspace{2px} ds \hspace{1px} dz \hspace{1px} dt} { 0.5 \hspace{1px } \rho_{\infty} U_{\infty}^2 L \hspace{1px} z_{span} } \mbox{ ,} \label{eq:cmu}
	\end{align}
	where $T_g$ is the period of $ G(t) $. Different values of $ C_{\mu} $ are tested to assess the effectiveness of flow control. Table \ref{table:controlsetup} displays all configurations investigated in terms of $ C_{\mu} $ for all actuator setups. For clarity, we will refer to simulations with a specific $ C_{\mu} $ as ``Case $ \# $''. In what follows, results are obtained for Case 2 at $ St = 5 $, unless otherwise stated.
	\renewcommand{\arraystretch}{1.2}	
	\setlength{\tabcolsep}{12pt}
	\begin{table}[]
		\caption{Parameters of control setups investigated. Simulations with the same $ C_{\mu} $ are grouped under the same Case category.}
		\vspace{-13px}
		\begin{tabular}{cccccc}
			\toprule \toprule
			\multirow{2}{*}{\textbf{Case}} & \multirow{2}{*}{\textbf{$C_{\mu}$}} & \multicolumn{4}{c}{\textbf{$ \frac{U_{jet \hspace{2px} max}}{U_{\infty}} $}}    \\ \cline{3-6} 
			&                                     & \textbf{2D Act.} & \textbf{2 slots (A)} & \textbf{2 slots (B)} & \textbf{3 slots (C)} \\ \midrule
			1                     & $ 1.78 \times 10^{-1} \hspace{5px} \% $          & 0.8              & -                    & -                    & -                \\ 
			2                     & $ 4.46 \times 10^{-2} \hspace{5px} \% $          & 0.4              & 0.90                 & 0.67                 & 0.74             \\
			3                     & $ 1.12 \times 10^{-2} \hspace{5px} \% $          & 0.2              & -                    & -                    & -                \\ \bottomrule \bottomrule

		\end{tabular}
		
		\label{table:controlsetup}
	\end{table}

\subsection{Flow Configuration and Mesh Convergence Study}

Large eddy simulations are performed for a SD7003 airfoil in a plunging motion described by Eq. \ref{eq:plunge} at Reynolds number $ Re = 60{,}000 $, freestream Mach number $ M = 0.1 $ and static angle of attack $\alpha_0 = 8^{\circ}$. The plunging motion has a reduced frequency $k = 0.5$ and the plunge amplitude is set as $ h_o = 0.5 L $. This specific flow condition was selected based on the availability of results from similar high fidelity simulations from \cite{Visbal:2011}. In this reference, simulations were performed for different spanwidths. It was concluded that the main flow features were fairly insensitive to spanwidth variations due to the energetic forcing of the plunging motion. Therefore, we employ a span length $z_{span} = 0.4 L$ in our calculations similarly to the baseline case from \cite{Visbal:2011}.

\begin{figure}[ht]
	\centering
	\begin{subfigure}{0.45\textwidth}
		\centering
		\includegraphics[width=\textwidth,trim={10px 10px 10px 10px}]{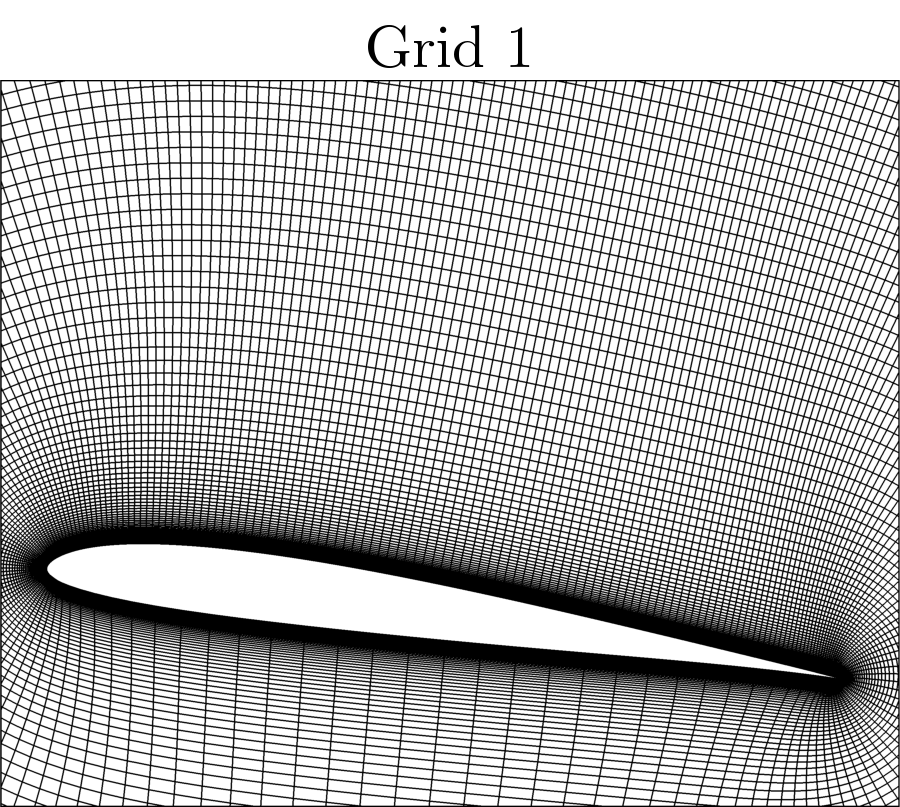}
	\end{subfigure}
	\hspace{10px}
	\begin{subfigure}{0.45\textwidth}
		\centering
		\includegraphics[width=\textwidth,trim={10px 10px 10px 10px}]{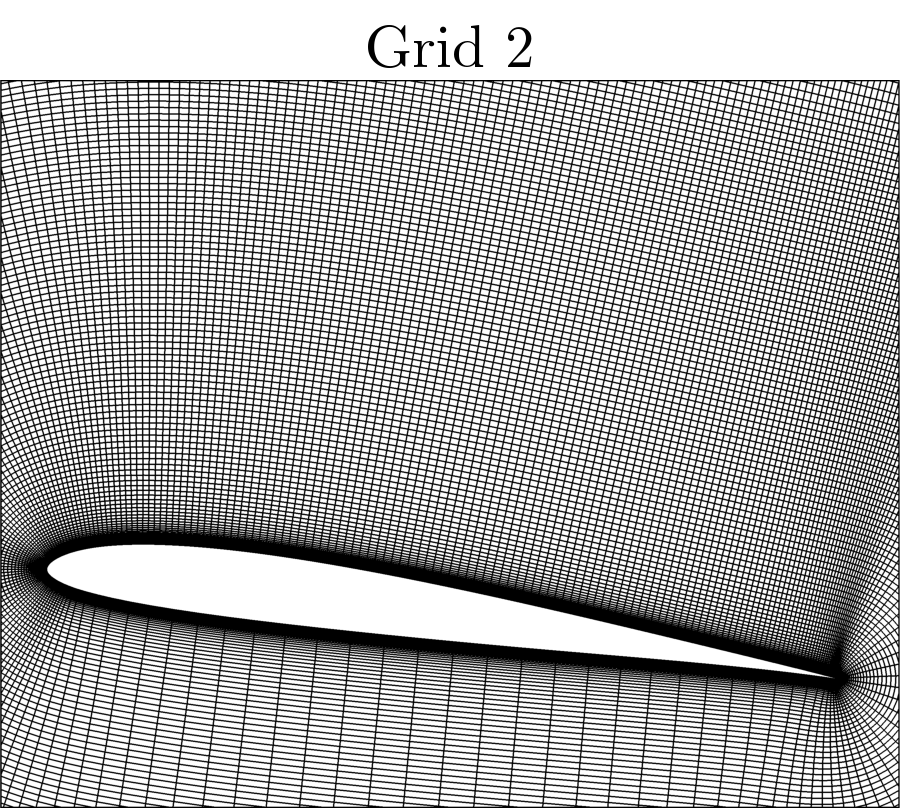}
	\end{subfigure}
	\caption{Grids considered in the mesh refinement study (only every other grid point in the $ x $-$ y  $ plane is shown here).}
	\label{fig:grid}
\end{figure}

A mesh convergence study is conducted to assess influence of grid resolution on the simulated flows.  Figure \ref{fig:grid} shows detail views of the two grids which are generated with approximately $70\%$ of the surface points located along the suction side of the airfoil. This setup is employed since turbulence appears in this region at various stages of the plunging motion and, hence, finer scales need to be resolved. At the pressure side, however, the flow does not become turbulent at any moment during the plunging motion. The trailing edge of the SD7003 airfoil is rounded in current simulations with an arc of radius $r/L = 0.0008 $. This procedure is required for maintaining the metric terms employed in the structured grid smooth. 

The grids parameters are listed in Table \ref{tab:grid}. In this study, we employ resolutions similar to those from Visbal \cite{Visbal:2011}. It is important to mention that a similar numerical approach was used by \cite{Visbal:2011} and, therefore, the current investigation follows the best practices needed to properly simulate the current flow. From grid 1 to 2, we mainly improved the spanwise resolution and the concentration of points in the wall-normal direction in the region comprised by a chord length to the airfoil surface. This latter refinement was achieved by changing the stretching function that defines the grid generation.
\renewcommand{\arraystretch}{1.2}	
\begin{table}[ht]
	\centering
	\caption{Grids Parameters. }
	\vspace{-15px}
	\label{tab:grid}
	\begin{tabular}{c c c c  c c }
		\toprule \toprule
		Grid & $\xi^1$ & $\xi^2$ & $\xi^3$ & ${\Delta \xi^2}_{wall}$ & ${\Delta \xi^2}^*$ \\ \midrule
		1 & 441   & 300   & 60    & 0.00005             & 0.01           \\ 
		2 & 481   & 350   & 96    & 0.00005             & 0.005          \\ \bottomrule \bottomrule
	\end{tabular} \par 
	\vspace{10px}
	\footnotesize ${\Delta \xi^2}_{wall}$: distance between airfoil surface and first grid point in the normal direction \par
	${\Delta \xi^2}^*$: distance between points in the normal direction one chord away from the airfoil
\end{table}

\begin{figure*}[ht]
		\centering
		\begin{subfigure}[t]{\3PerLine \textwidth}
			\centering
			\includegraphics[width = 1\textwidth]{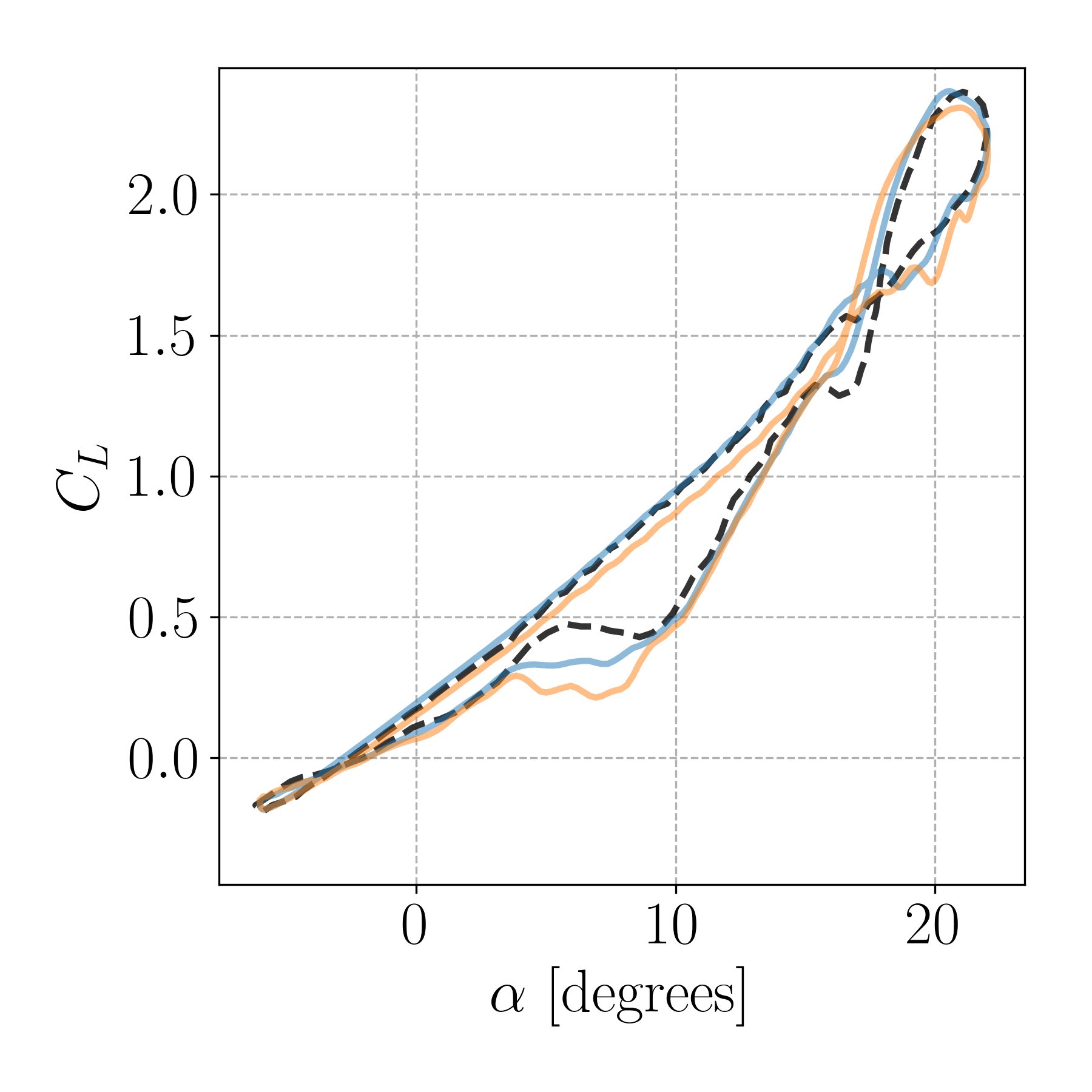}
		\end{subfigure}
		\begin{subfigure}[t]{\3PerLine \textwidth}
			\centering
			\includegraphics[width = 1\textwidth]{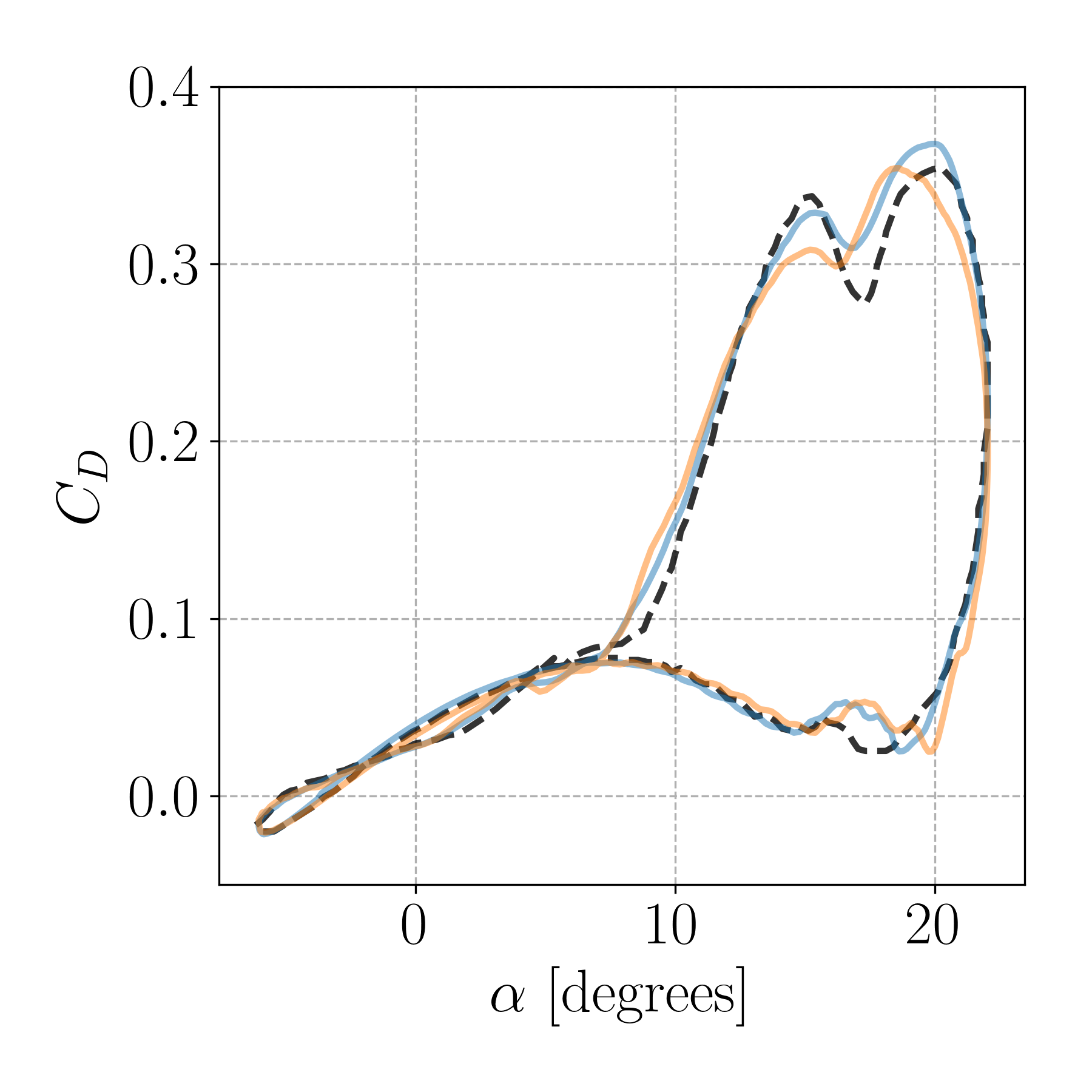}
		\end{subfigure}
		\begin{subfigure}[t]{\3PerLine \textwidth}
			\centering
			\includegraphics[width = 1\textwidth]{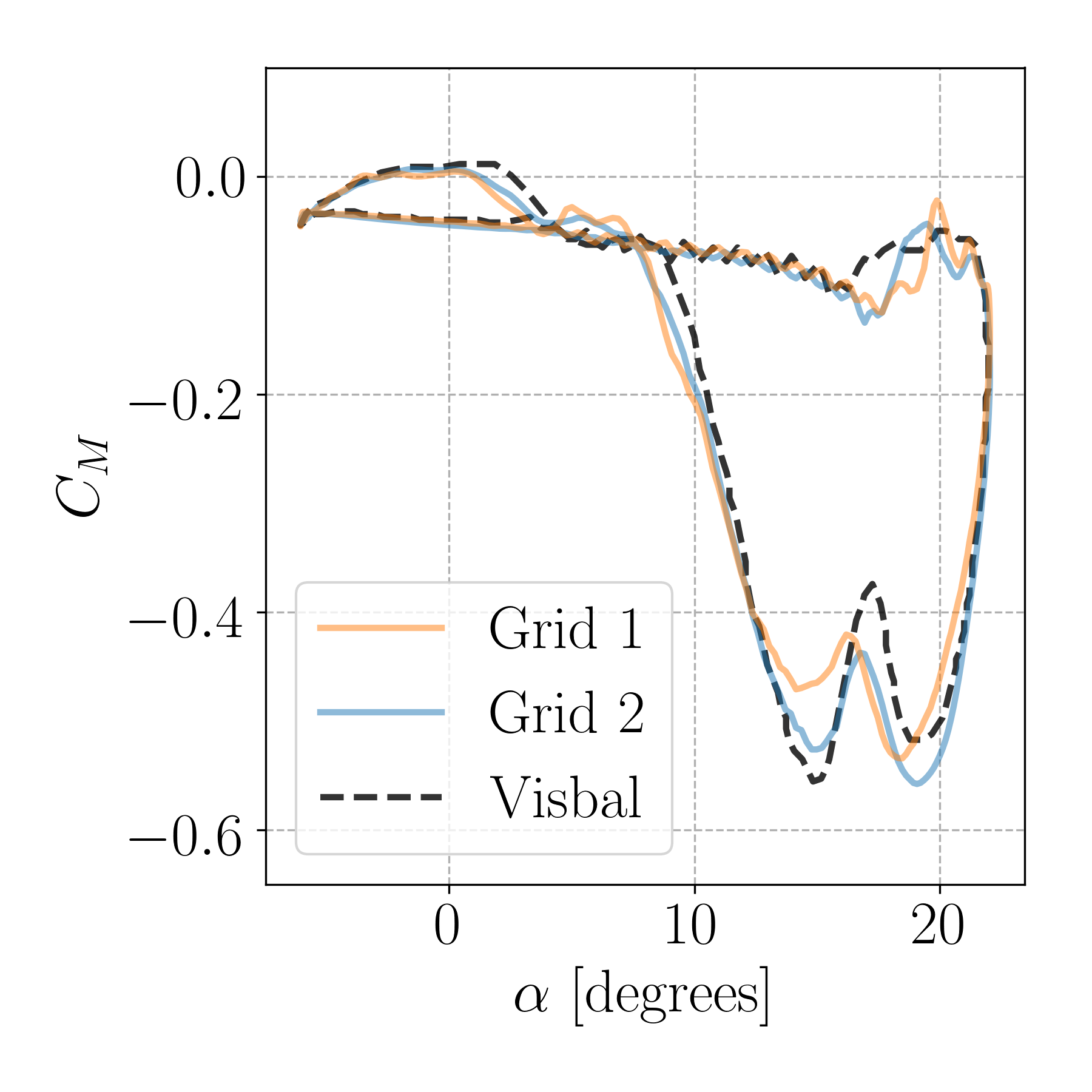}
		\end{subfigure}
		\caption{Aerodynamic coefficients obtained using grids 1 and 2 and from Ref. \cite{Visbal:2011} as function of the effective angle of attack $ \alpha $. }
		\label{fig:grids_clcdcm}
	\end{figure*}

\begin{figure*}[ht]
		\centering
		\begin{subfigure}[b]{0.32 \textwidth}
			\centering
			\includegraphics[width=1\textwidth]{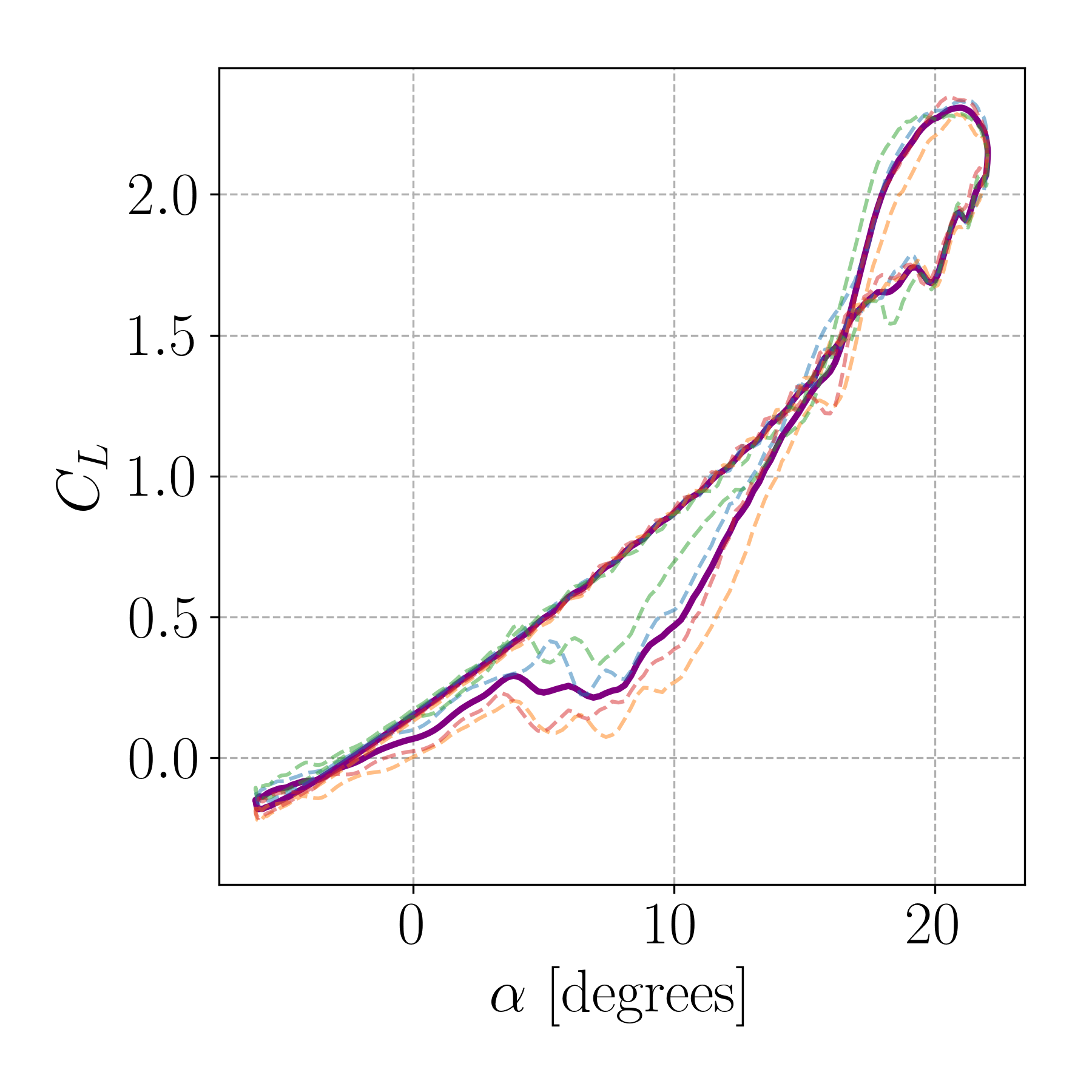}
		\end{subfigure}
		%
		\begin{subfigure}[b]{0.32 \textwidth}
			\centering
			\includegraphics[width=1\textwidth]{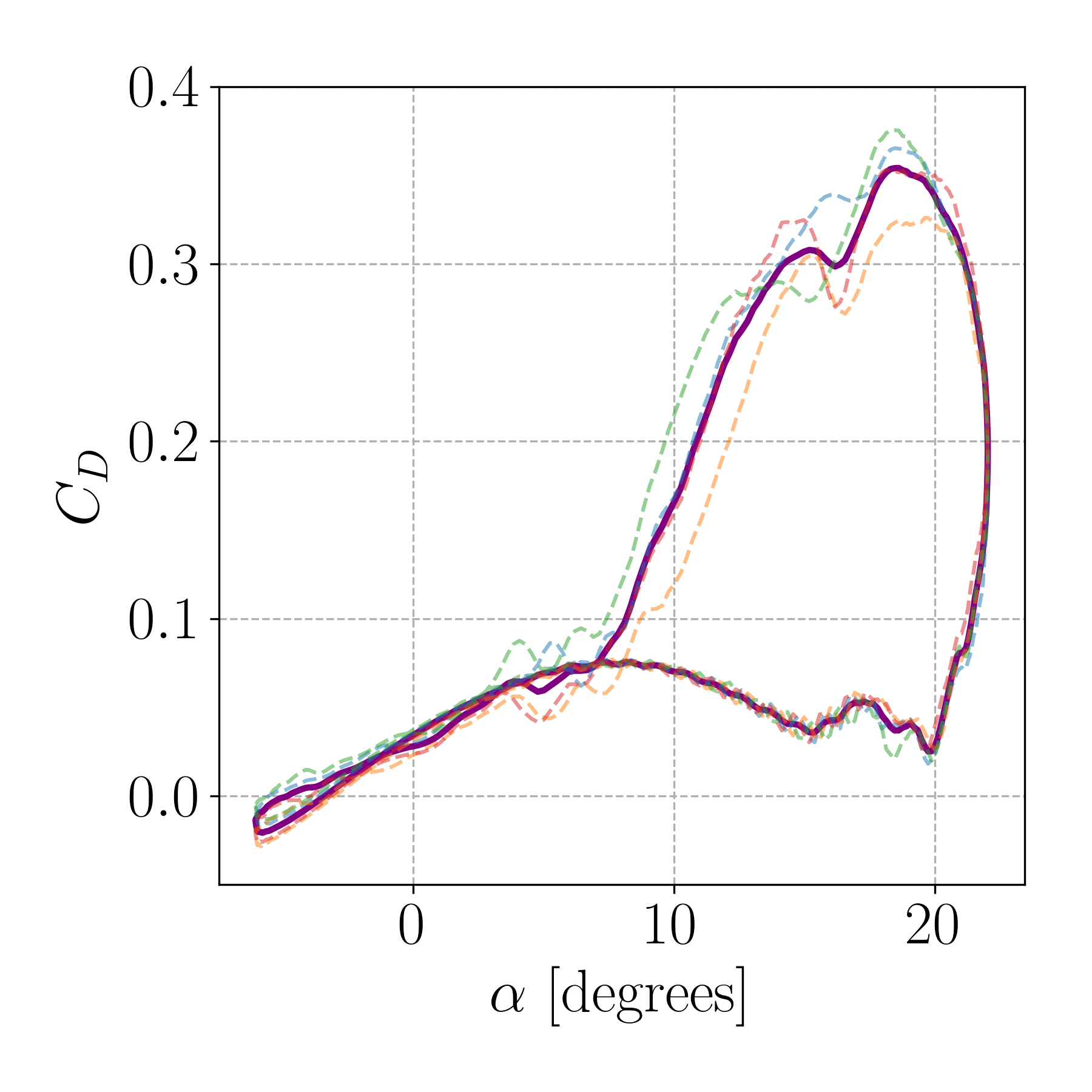}
		\end{subfigure}
		%
		\begin{subfigure}[b]{0.32 \textwidth}
			\centering
			\includegraphics[width=1\textwidth]{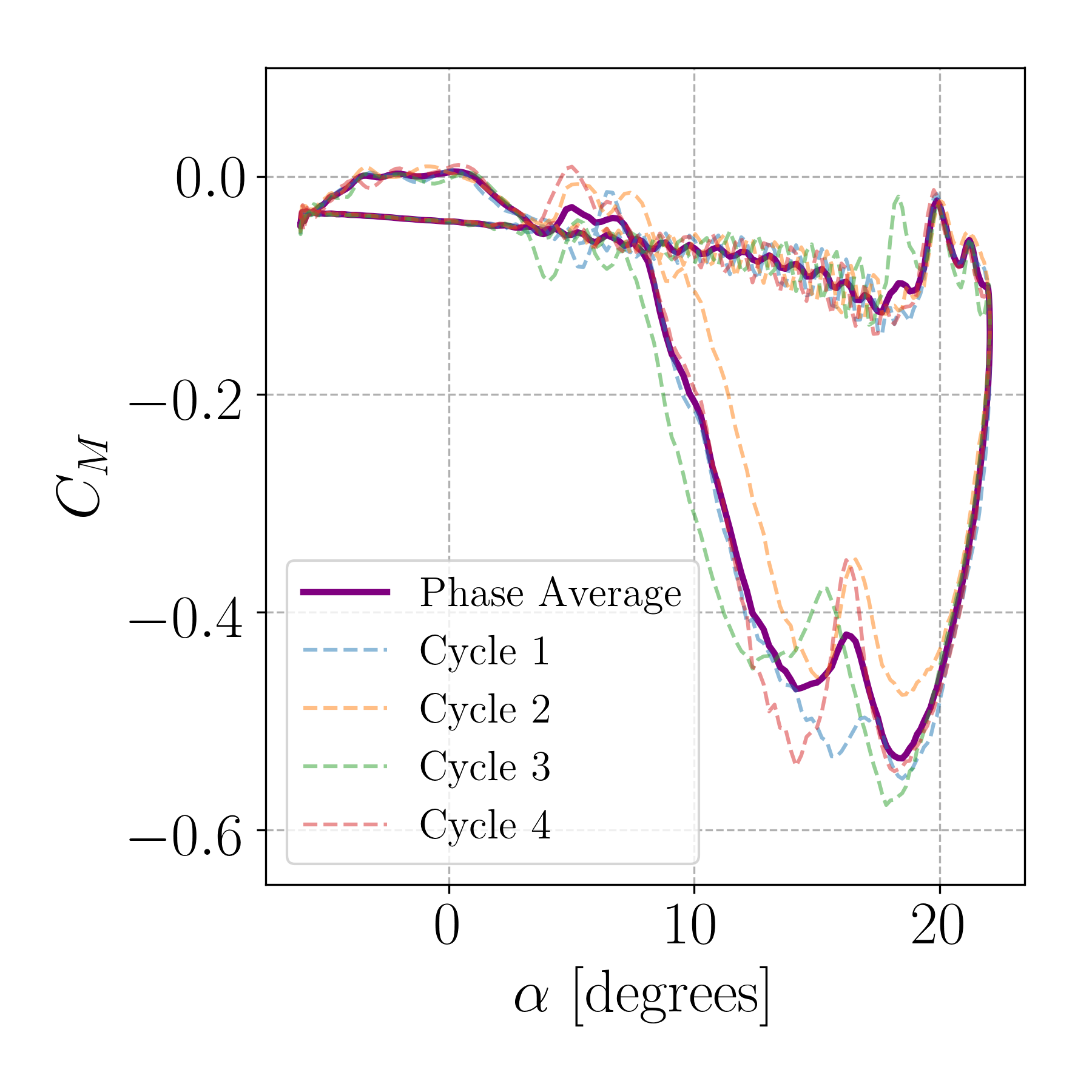}
		\end{subfigure}		
		\caption{Cycle to cycle variations in aerodynamic coefficients (grid 1).}	
		\label{fig:cxs}	
	\end{figure*}

	
Simulations of five cycles of plunging motion are performed, but only the last four are used to calculate the phase-averaged statistics. Figure \ref{fig:grids_clcdcm} shows the phase-averaged lift, drag and quarter-chord pitching moment coefficients, $ C_L $, $ C_D $ and $ C_M $, respectively, with respect to the effective angle of attack $ \alpha = \alpha_0 + \tan^{-1}\left( \frac{k h_0}{L} \cos ( k t ) \right) $.
	\def\3PerLine {0.32}
	Results obtained using both grids exhibit good agreement with \cite{Visbal:2011}, especially considering the variations that occur from cycle to cycle. Such variations can be seen in Fig. \ref{fig:cxs}, in which aerodynamic coefficients obtained by the first cycle are already discarded and only the last four are employed in computations. From current results, we consider that the coarser mesh shown in Fig. \ref{fig:grid} has sufficient resolution to capture the flow physics. Hence, this mesh is chosen to perform the simulations presented in this work.

\section{Flow Features of Baseline Configuration}
	
\begin{figure}[ht]
		\centering
		\includegraphics[width=0.45\textwidth,trim={0 10px 0 0}, clip]{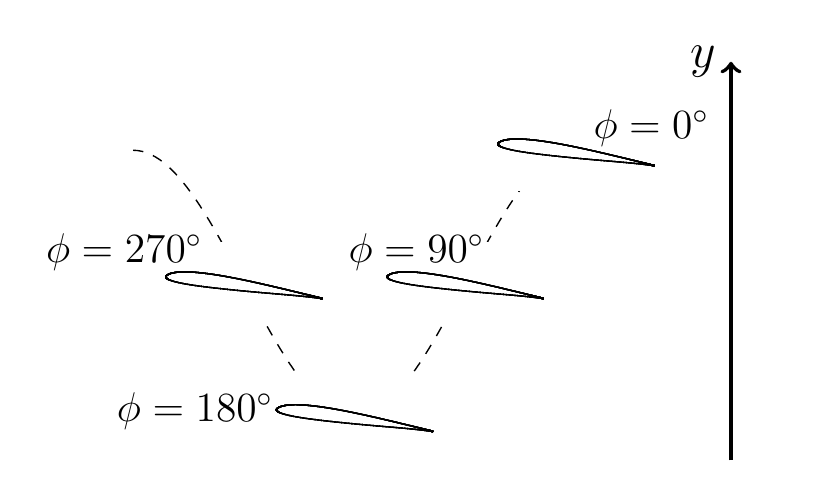}
		\caption{Airfoil position for different phase angles $\phi$.}
		\label{fig:phase_angles}	
	\end{figure}

	This section presents results of the current ILES for the baseline uncontrolled configuration, in which the main physical mechanisms associated with the dynamic stall vortex are described. The current plunge motion undergoes an effective angle of attack in the range of $-6^{\circ}$ $ \leq \alpha \leq 22^{\circ}$. Due to transients originated from the start of the simulations, only the last four plunging cycles from all five available are used to calculate statistics. For visualization purposes, a phase angle $\phi$ is used to describe the airfoil position. A schematic of the airfoil motion is shown in Fig. \ref{fig:phase_angles}. At $\phi=0^{\circ}$, the airfoil has no vertical velocity and is at the top-most position of the plunging motion. At $\phi=90^{\circ}$, it has the highest downward velocity in the $y$-direction and, at $\phi=180^{\circ}$, it has zero vertical velocity being at the bottom-most position of the plunging motion. Finally, at $\phi=270^{\circ}$ it has the highest velocity in the $y$-direction (upward). 
	\def\3PerLine{0.32}
	\begin{figure*}[ht!]
		\centering
		\begin{subfigure}[t]{\3PerLine \textwidth}
			\centering
			\includegraphics[width=\textwidth]{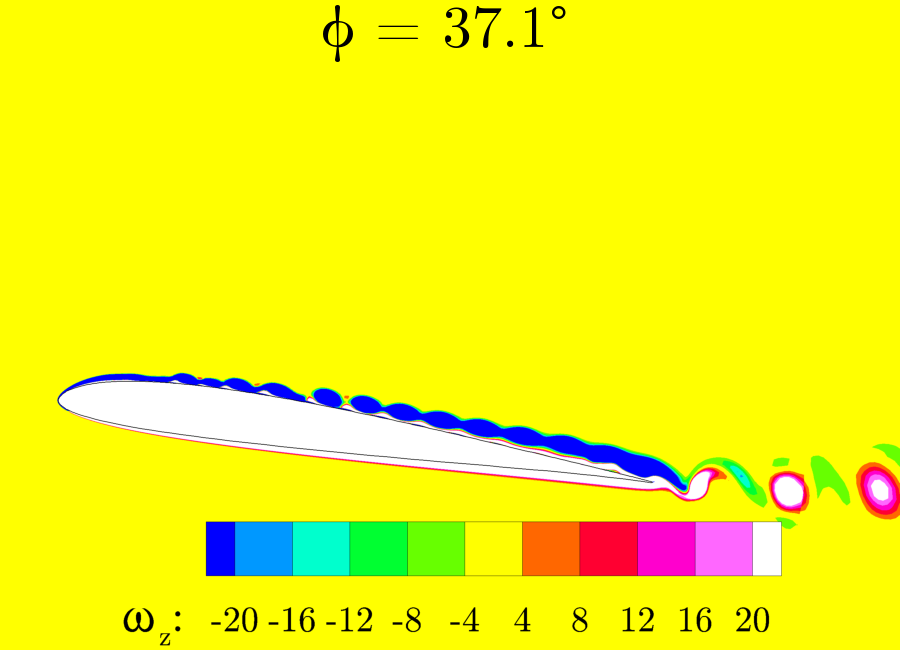}
			\subcaption{}
			\label{fig:inst}
		\end{subfigure}	
		\begin{subfigure}[t]{\3PerLine \textwidth}
			\centering
			\includegraphics[width=\textwidth]{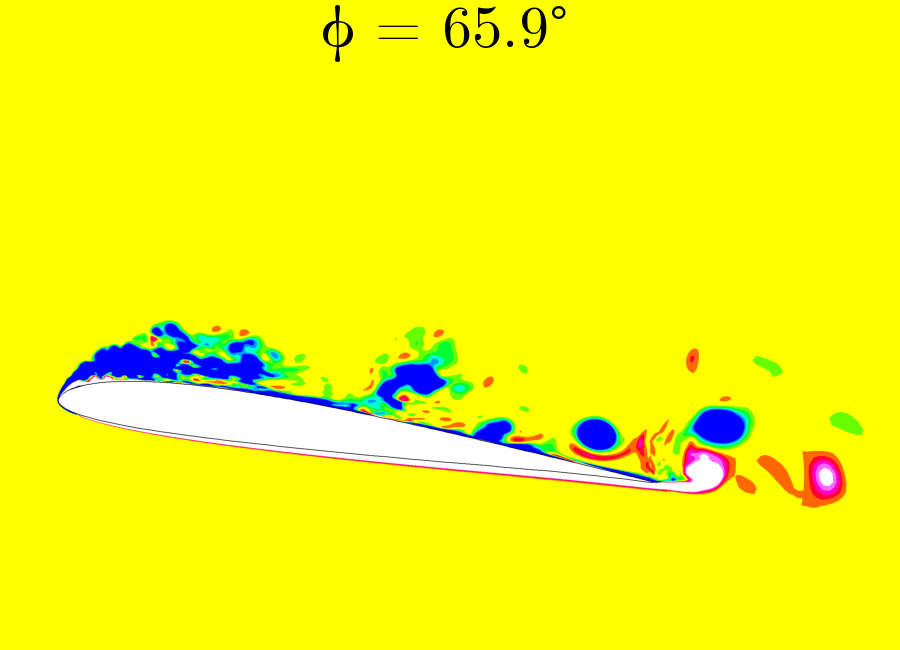}
			\subcaption{}
			\label{fig:LEV_formation}
		\end{subfigure}	
		
		\begin{subfigure}[t]{\3PerLine \textwidth}
			\centering
			\includegraphics[width=\textwidth]{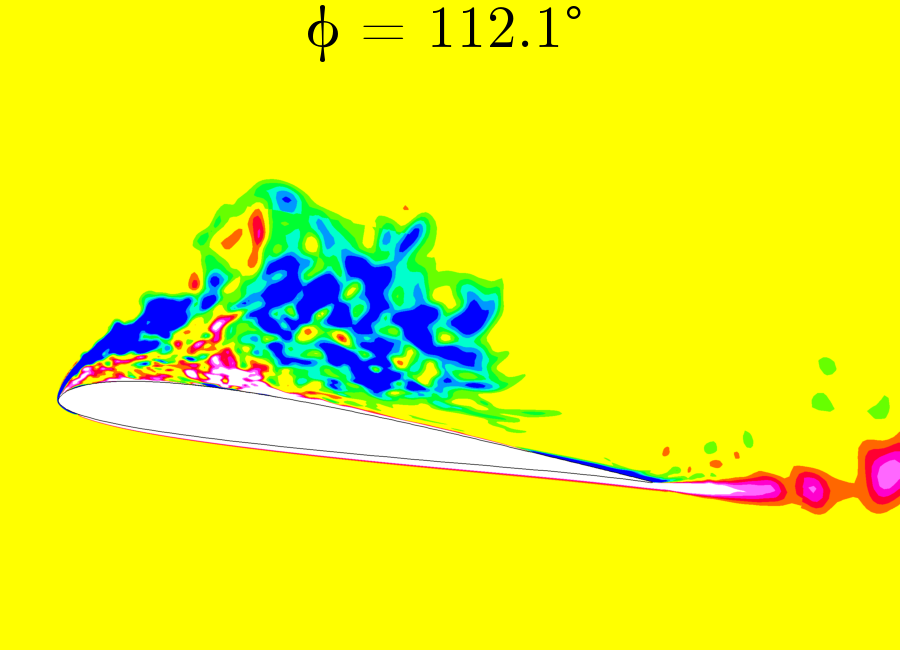}
			\subcaption{}
			\label{fig:LEV_convection}
		\end{subfigure}	
		\begin{subfigure}[t]{\3PerLine \textwidth}
			\centering
			\includegraphics[width=\textwidth]{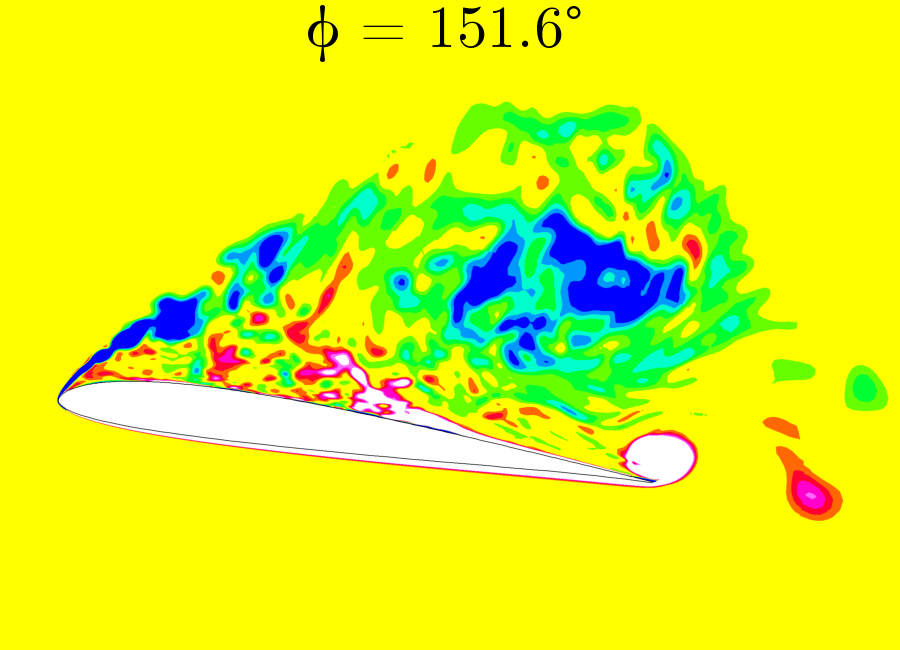}
			\subcaption{}
			\label{fig:TEV_push}
		\end{subfigure}		
		\begin{subfigure}[t]{\3PerLine \textwidth}
			\centering
			\includegraphics[width=\textwidth]{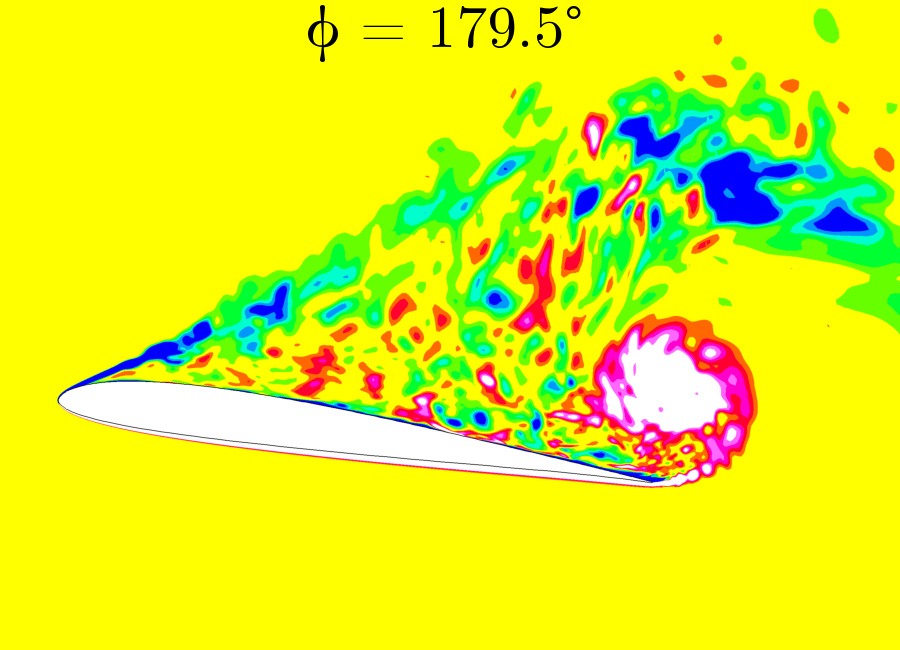}
			\subcaption{}
			\label{fig:begin_laminar}
		\end{subfigure}	
		\begin{subfigure}[t]{\3PerLine \textwidth}
			\centering
			\includegraphics[width=\textwidth]{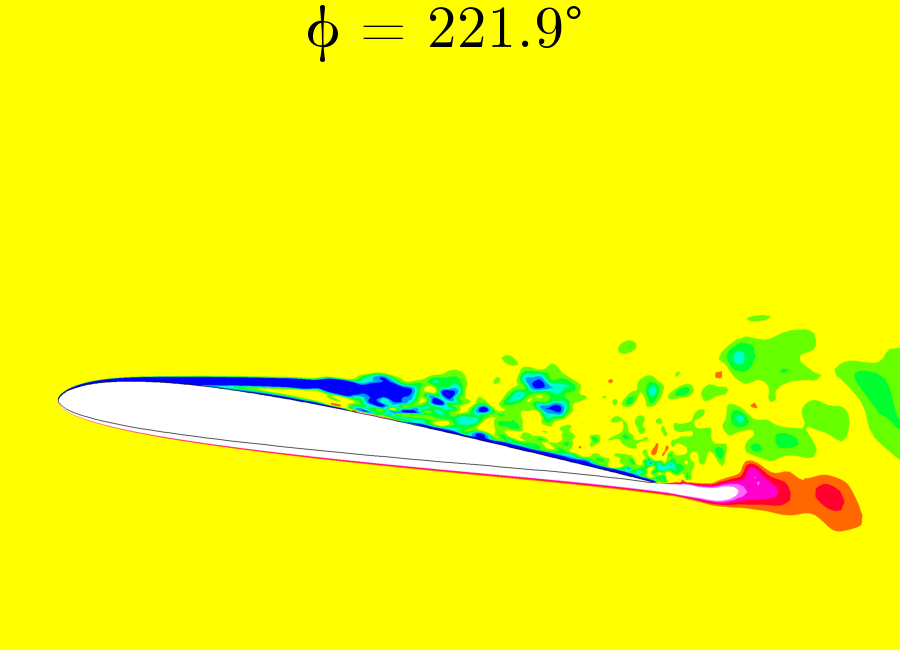}
			\subcaption{}
			\label{fig:mid_laminar}
		\end{subfigure}	
		\begin{subfigure}[t]{\3PerLine \textwidth}
			\centering
			\includegraphics[width=\textwidth]{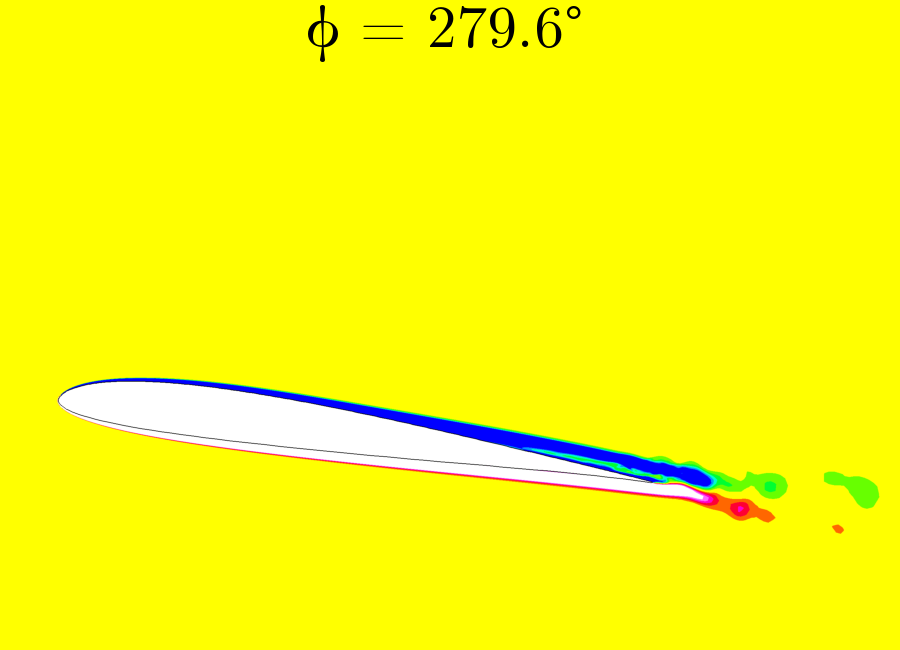}
			\subcaption{}
			\label{fig:end_laminar}
		\end{subfigure}	
		\caption{Spanwise-averaged vorticity contours at different phases of the plunging motion without actuation (see also Supplemental Material \cite{sup1}).}
		\label{fig:z_vorticity}
	\end{figure*}
		
	Figure \ref{fig:z_vorticity} and Supplemental Material \cite{sup1} present spanwise-averaged vorticity contours at different phases of the plunging cycle. During the downstroke, flow instabilities begin to grow in the shear layer formed along the suction side of the airfoil with vortex shedding occurring at the airfoil wake as shown in Fig. \ref{fig:inst}. As the downward motion continues, instabilities on the suction side grow and eventually break the large spanwise-correlated structures into finer ones, leading to a transitional flow. While this takes place, the main leading-edge vortex (LEV) begins to form as shown in Fig. \ref{fig:LEV_formation}. The LEV grows over the suction side (Fig. \ref{fig:LEV_convection}), increasing lift and creating a nose-down pitching moment. 

	As the LEV covers the entirety of the chord, a trailing-edge vortex (TEV) forms and ``lifts" the LEV away from the airfoil surface, as shown in Fig. \ref{fig:TEV_push}. As the LEV lifts off, an oscillation in the pitching moment can be observed. As the airfoil motion continues, the TEV is ejected from the suction side (Fig. \ref{fig:begin_laminar}). When the airfoil moves upwards, re-laminarization starts from the leading edge (Fig. \ref{fig:mid_laminar}) and keeps going until the entire boundary layer is relaminarized (Fig. \ref{fig:end_laminar}). Subsequently, the Kelvin--Helmholtz instability can be observed again, leading to periodic shedding of vortices from the trailing edge.
	\begin{figure*}[ht]
		\centering
		\includegraphics[width=\textwidth]{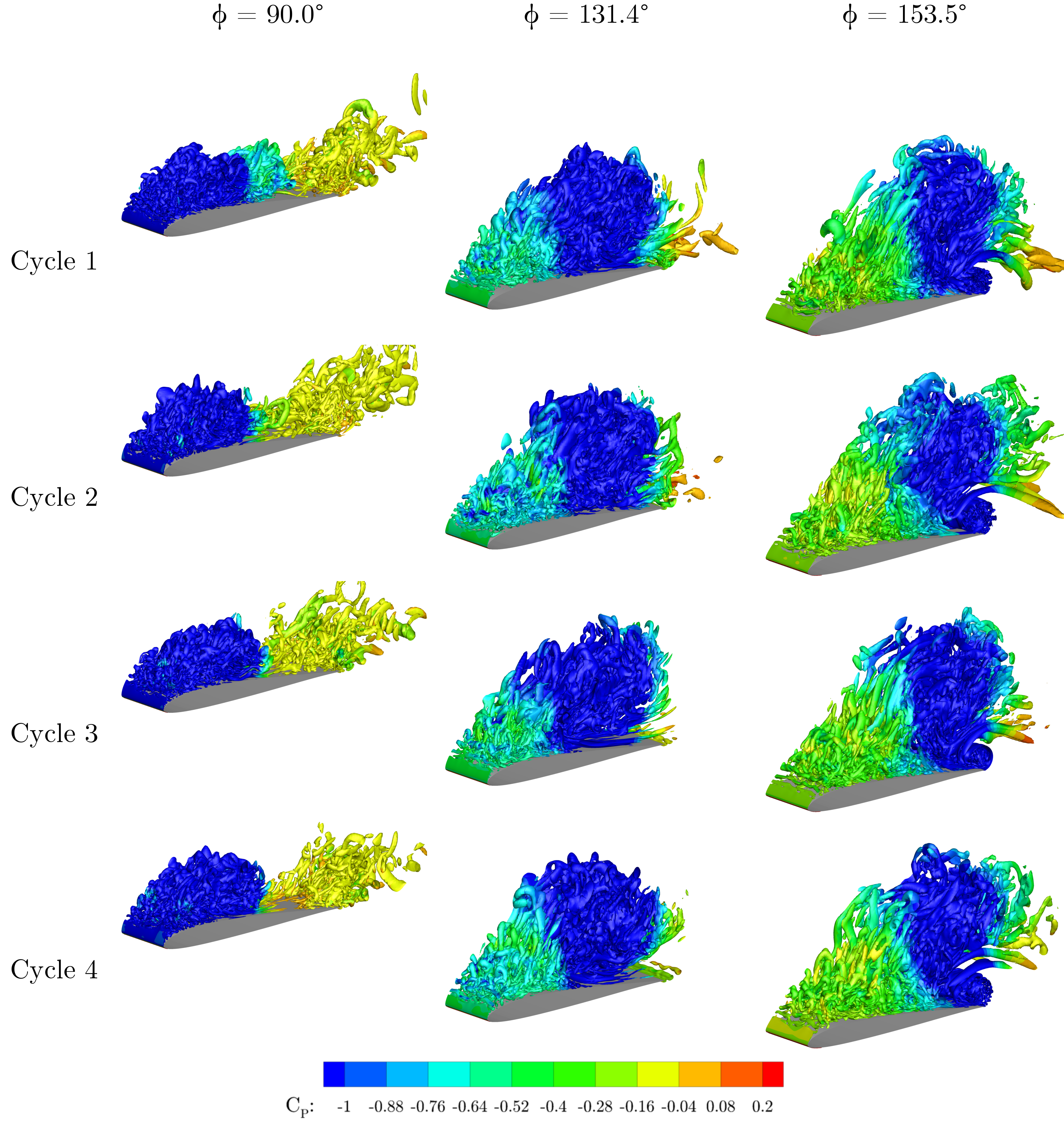}
		\caption{Iso-surfaces of Q criterion colored by $ C_P $ at different phases of the plunge motion without actuation.}
		\label{fig:q_criterion}
	\end{figure*}
	
	In order to further characterize the current flow, iso-surfaces of Q-criterion are shown in Fig. \ref{fig:q_criterion} for all cycles. The turbulent structures are colored by pressure coefficient contours. Despite subtle cycle to cycle variations, the main features of the dynamic stall process remain unchanged. Namely, the formation of the LEV, its transport over the airfoil, the formation of the TEV and the departure of both vortices. Although fine turbulent structures can be observed, it is clear that large-scale coherent structures are the most prominent in the dynamic stall process. We expect such energetic structures to play a key role in the dynamics of the present flow, severely impacting the   aerodynamic loads. For example, the leading-edge vortex is characterized by a low pressure region which is advected along the suction side, dynamically affecting flight stability through changes in lift and drag forces during the plunging motion. The next section describes the efforts towards controlling the formation of these structures to reduce overall drag and its fluctuations, while keeping lift unaltered.

\section{Active Flow Control}
	
\subsection{2D Actuation}

	An assessment of 2D actuation on the flow dynamics is presented in this section. We present the control effect based on a single cycle evaluation. Flow actuation is turned on at $ \phi = 0^{\circ}$ after five plunging cycles. Figure \ref{fig:comparison_control_cases} shows  the averaged values of $ C_L$, $ C_D $ and $ C_M $ represented by black dots for different actuation frequencies $St$. The maximum and minimum values of the aerodynamic coefficients computed during the cycle are given by the top and bottom values of each bar. Results obtained for the baseline configuration are depicted by orange bars while green, blue and red bars represent solutions computed for cases 1, 2 and 3, respectively, as described in Table \ref{table:controlsetup}. It is important to remind that the coefficient of momentum $C_{\mu}$ for case 1 is the highest investigated while that for case 3 is the lowest. Hence, this figure allows an assessment of the effects of 2D actuation in terms of both actuation frequency and its intensity on the aerodynamic coefficients.
	\def \temp{0.45}
	\begin{figure*}[ht]
		\centering
		\begin{subfigure}[t]{\temp \textwidth}
			\centering
			\includegraphics[width=1\textwidth, trim={20px 30px 20px 20px}, clip]{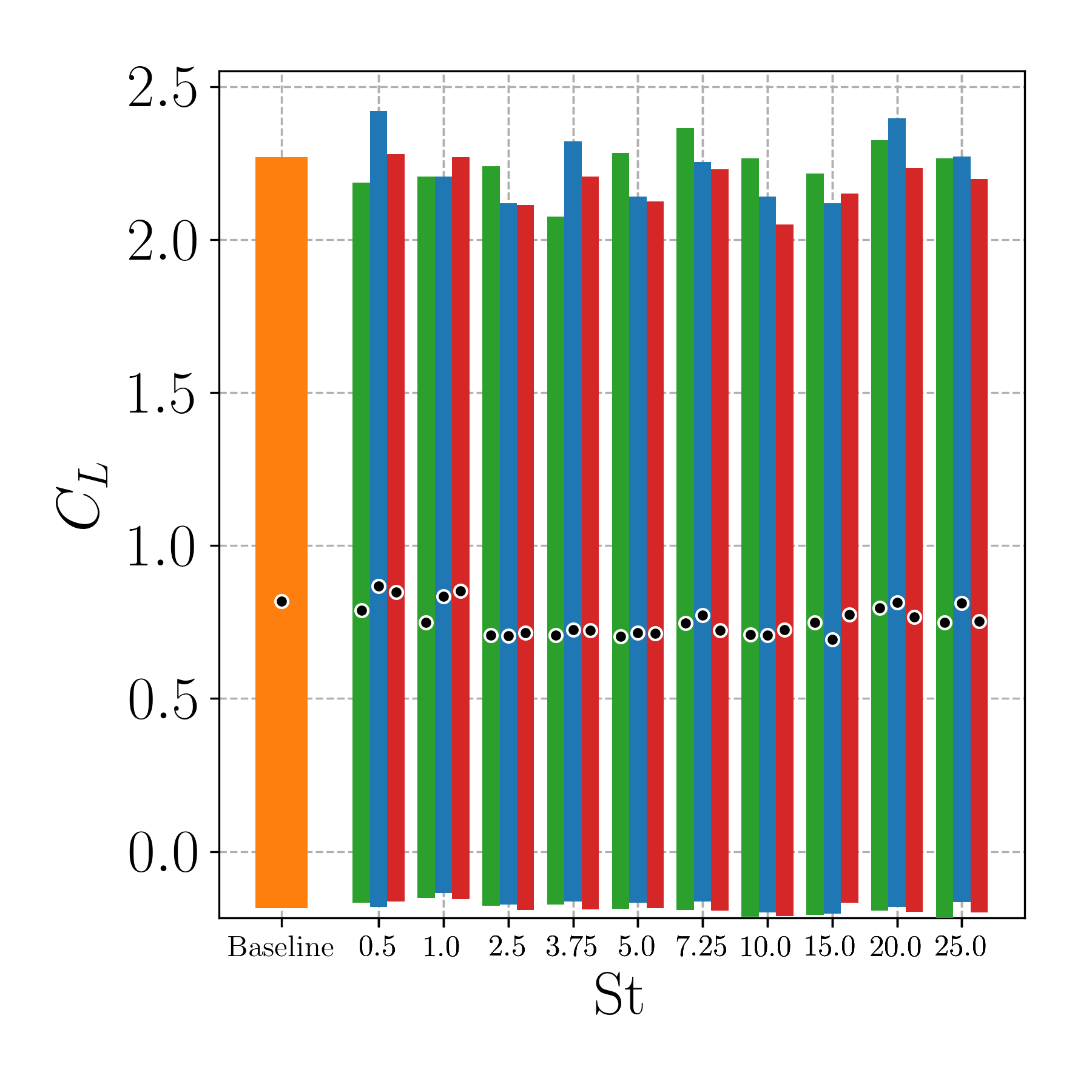}		
			\label{control_stats_cl}	
		\end{subfigure}
		\begin{subfigure}[t]{\temp \textwidth}
			\centering
			\includegraphics[width=1\textwidth, trim={20px 30px 20px 20px}, clip]{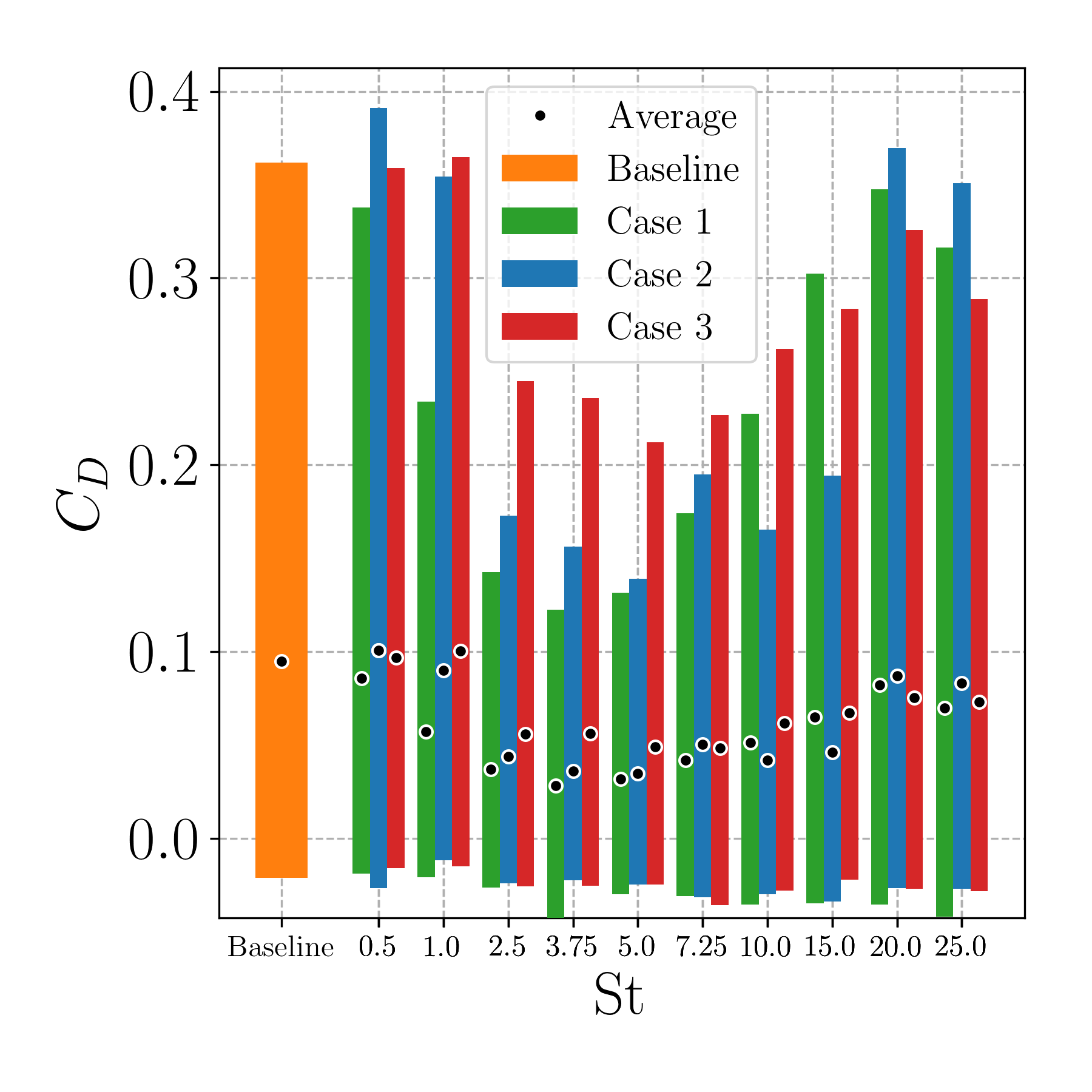}	
			\label{control_stats_cd}		
		\end{subfigure}				
		\begin{subfigure}[t]{\temp \textwidth}
			\centering
			\includegraphics[width=1\textwidth, trim={20px 30px 20px 20px}, clip]{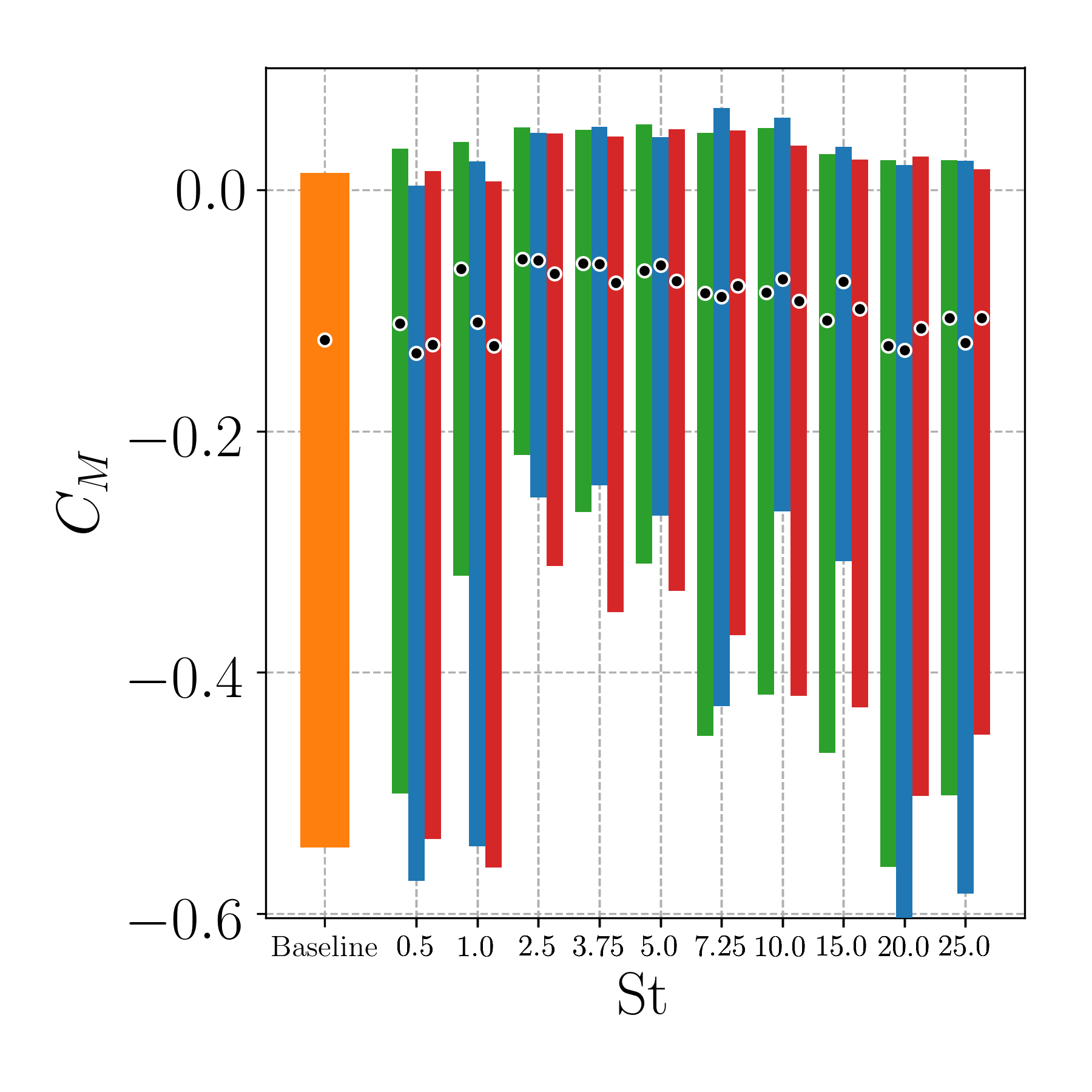}		
		\end{subfigure}
		\caption{Variations in aerodynamic coefficients for different actuation frequencies ($St$) and coefficient of momentum ($C_{\mu}$) for 2D actuated flows. We refer to simulations with different $ C_{\mu} $ as ``Case $ \# $''.}
		\label{fig:comparison_control_cases}
	\end{figure*}
	
	From Fig. \ref{fig:comparison_control_cases}, it can be noticed that $ C_L $ do not exhibit large variations for the actuation frequencies  and $ C_{\mu} $ considered. However, significant changes in $ C_D $ and $ C_M $ are observed depending on the actuation frequency. For example, large reductions in $ C_D $ appear in the range $ 2.5 < St < 15 $ compared to the baseline case for all values of $ C_{\mu} $ investigated. Frequencies higher than $ St = 15$ or lower than $ St = 2.5$ do not promote a significant impact on drag and pitching moment, both in terms of mean values and maximum and minimum amplitudes. The coefficient of momentum also has a significant impact on the results. In general, for the flows with stronger actuation disturbances (cases 1 and 2), reductions in maximum drag are more evident. In some occasions, better results in terms of drag reduction are observed for Case 2. 
		\begin{figure}
			\centering
			\begin{subfigure}{0.45\textwidth}
				\centering
				\includegraphics[width=\textwidth]{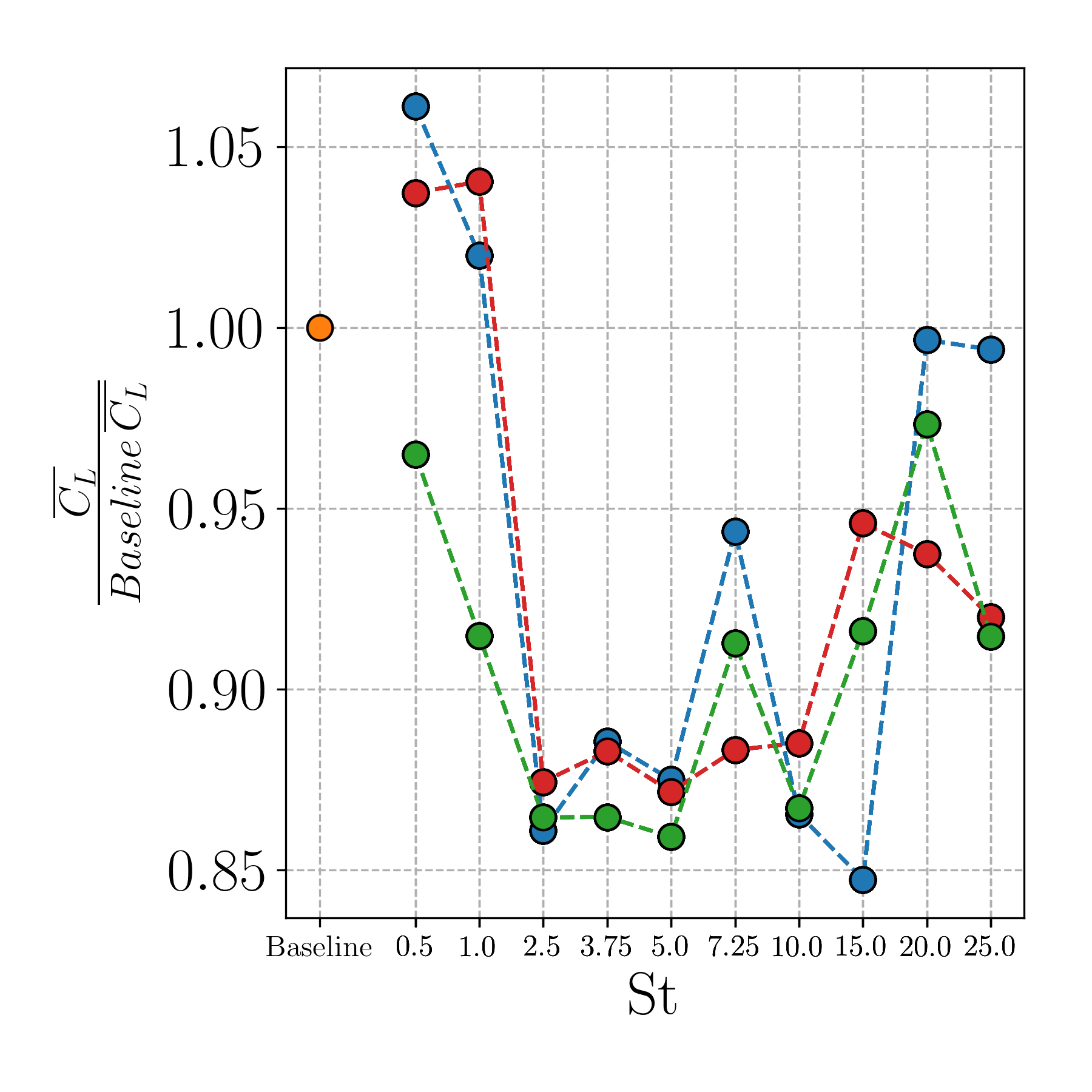}
			\end{subfigure}
			\begin{subfigure}{0.45\textwidth}
				\centering
				\includegraphics[width=\textwidth]{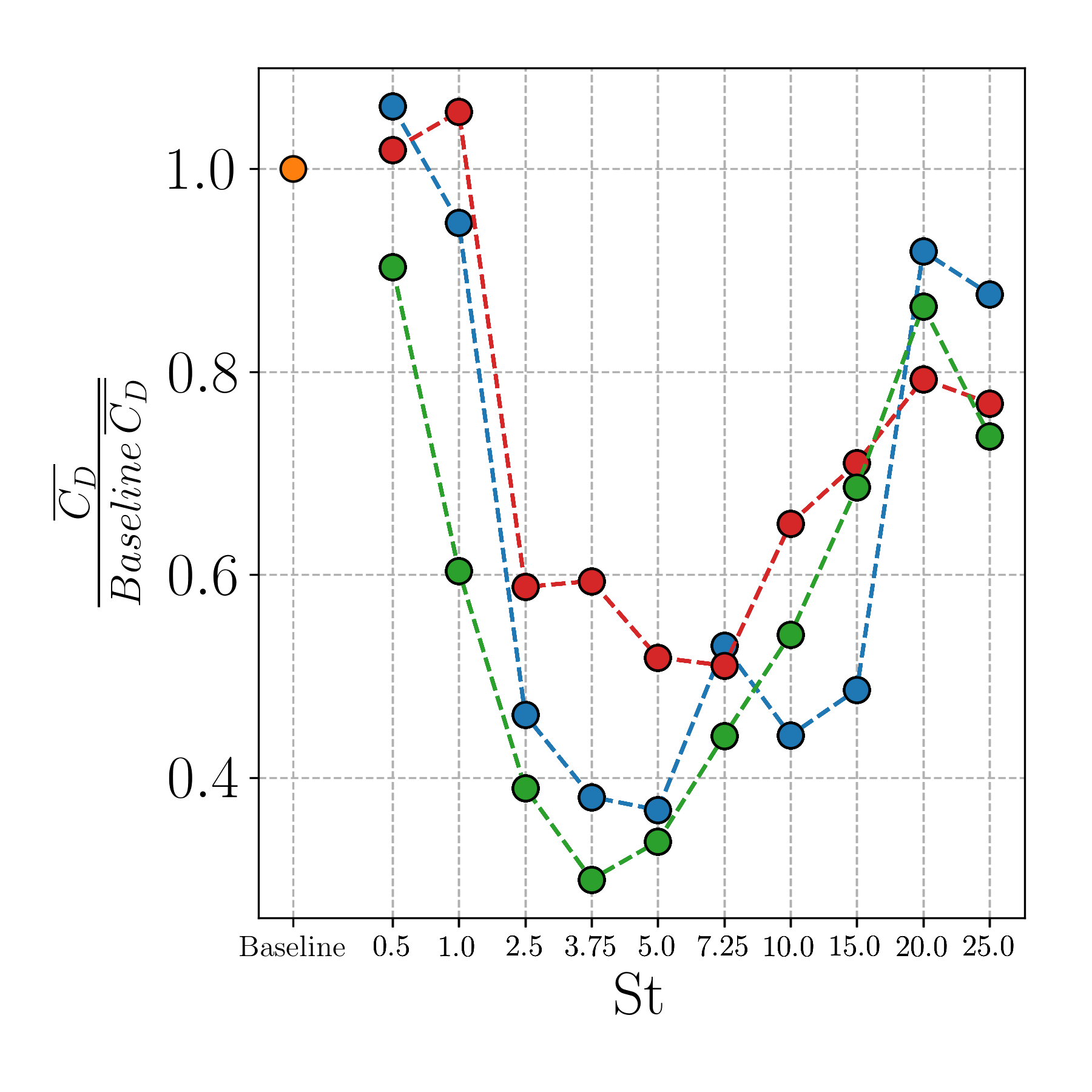}
			\end{subfigure}
			
			\begin{subfigure}{0.45\textwidth}
				\centering
				\includegraphics[width=\textwidth]{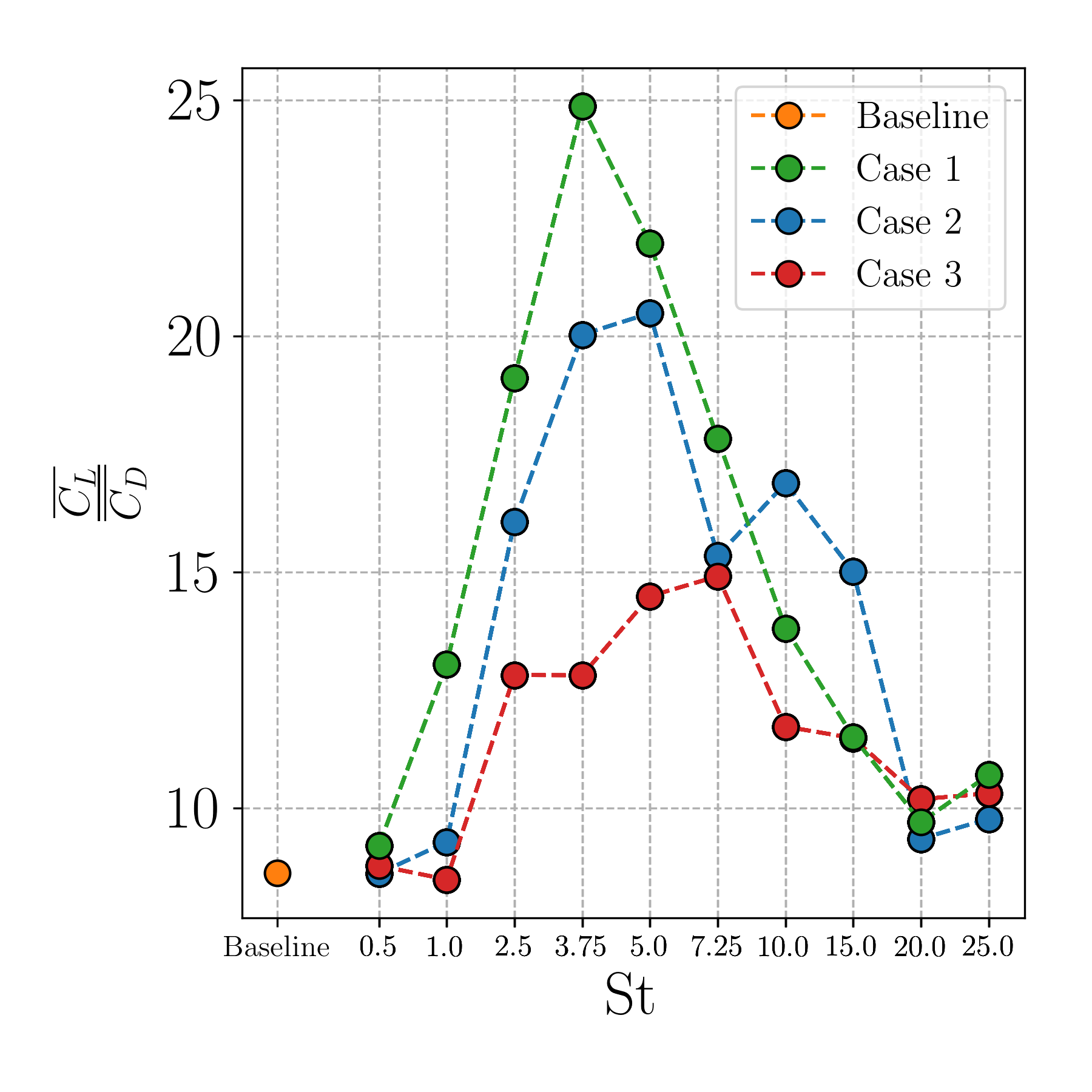}
			\end{subfigure}		
			\begin{subfigure}{0.45\textwidth}
				\centering
				\includegraphics[width=\textwidth]{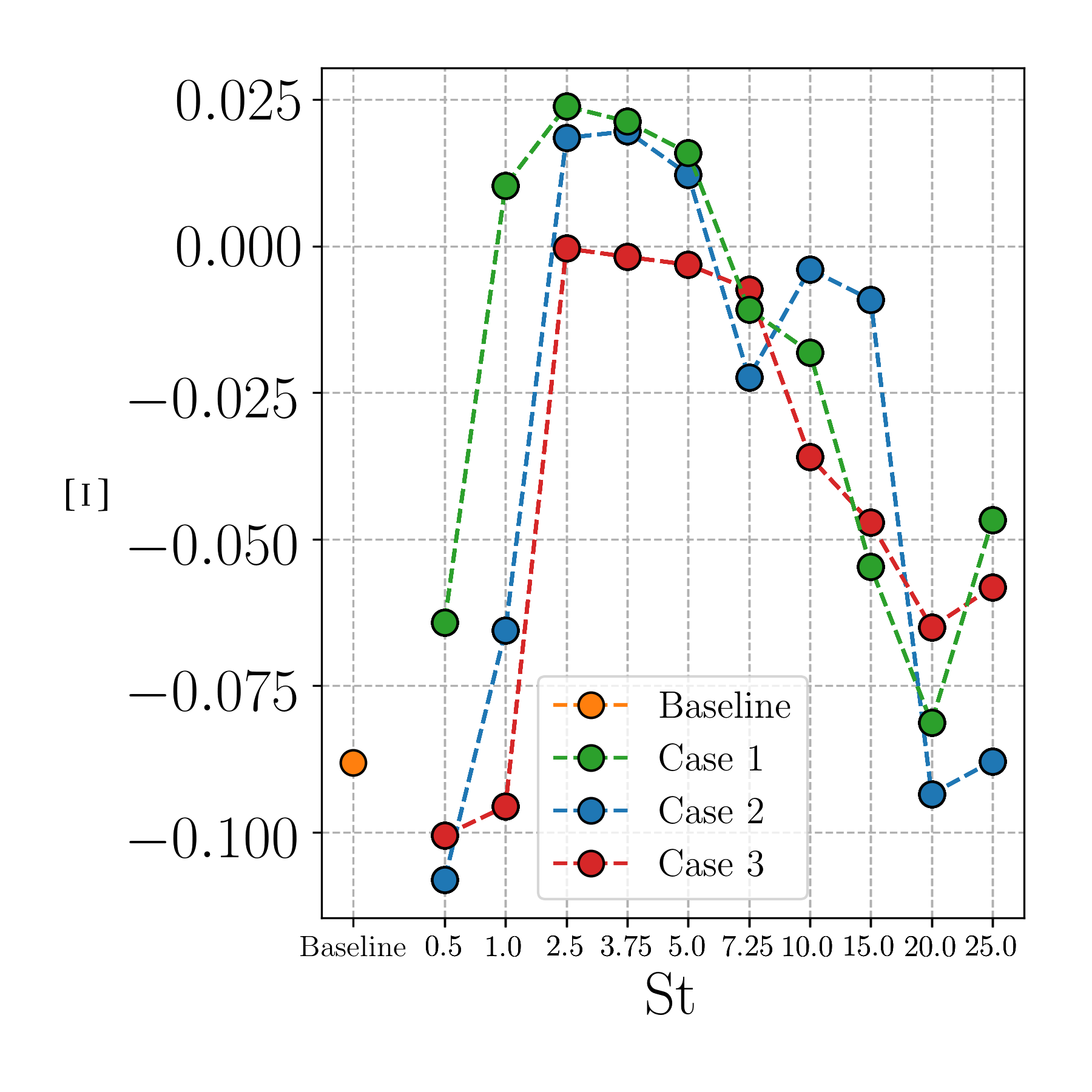}
			\end{subfigure}					
			\caption{Mean aerodynamic loads compared to the baseline flow, mean lift to mean drag ratio, and aerodynamic damping using 2D actuation.}
			\label{fig:statistics2DAct}
		\end{figure}
		
	Averaged values of $ C_L$ and $C_D$ normalized by their respective baseline values  are displayed in Fig. \ref{fig:statistics2DAct}. This figure also shows a drag polar plot relating $ \frac{\overline{C_L}}{\overline{C_D}} $. Again, results are presented as a function of Strouhal number and coefficient of momentum. The behavior observed for the maximum and minimum values of aerodynamic coefficients is similar to their averaged values. For example, with $ St = 3.75 $ and $C_{\mu}$ from case 1, the airfoil drag coefficient $ C_D $ is reduced to 30\% of the baseline. For the same case, the lift coefficient $ C_L $ only drops to 86\% of the baseline. In summary, for cases 1 and 2 and Strouhal numbers in the range $2.5 \leq St \leq 15$, flow actuation is able to considerably reduce mean values of drag coefficient without severely impacting lift. From the figure, one can conclude that the best results in terms of mean lift to mean drag ratio are obtained for frequencies given by $St = 3.75$ and $5.0$. 
	
	Figure \ref{fig:statistics2DAct} also shows the impact of actuation in the aerodynamic damping $ \Xi $, which is calculated as 
	\begin{equation}
    \Xi = -\frac{1}{\alpha_{max}-\alpha_0} \oint C_M \mathrm{d}\alpha \mbox{ .}
	\end{equation}
	It can be seen that the baseline flow has negative damping, implying that energy is  transferred from the flow to the airfoil, leading to oscillations and even flutter. While some actuation frequencies, e.g. cases 2 and 3 at $ St = 0.5 $, lead to even more negative values of aerodynamic damping, frequencies around $ St = 3.75 $ successfully revert the issue, leading to a positive damping and a stabilizing effect on the airfoil dynamics. 
	
\begin{figure*}[ht]
		\centering
		\begin{subfigure}[b]{0.32 \textwidth}
			\centering
			\includegraphics[width=1\textwidth]{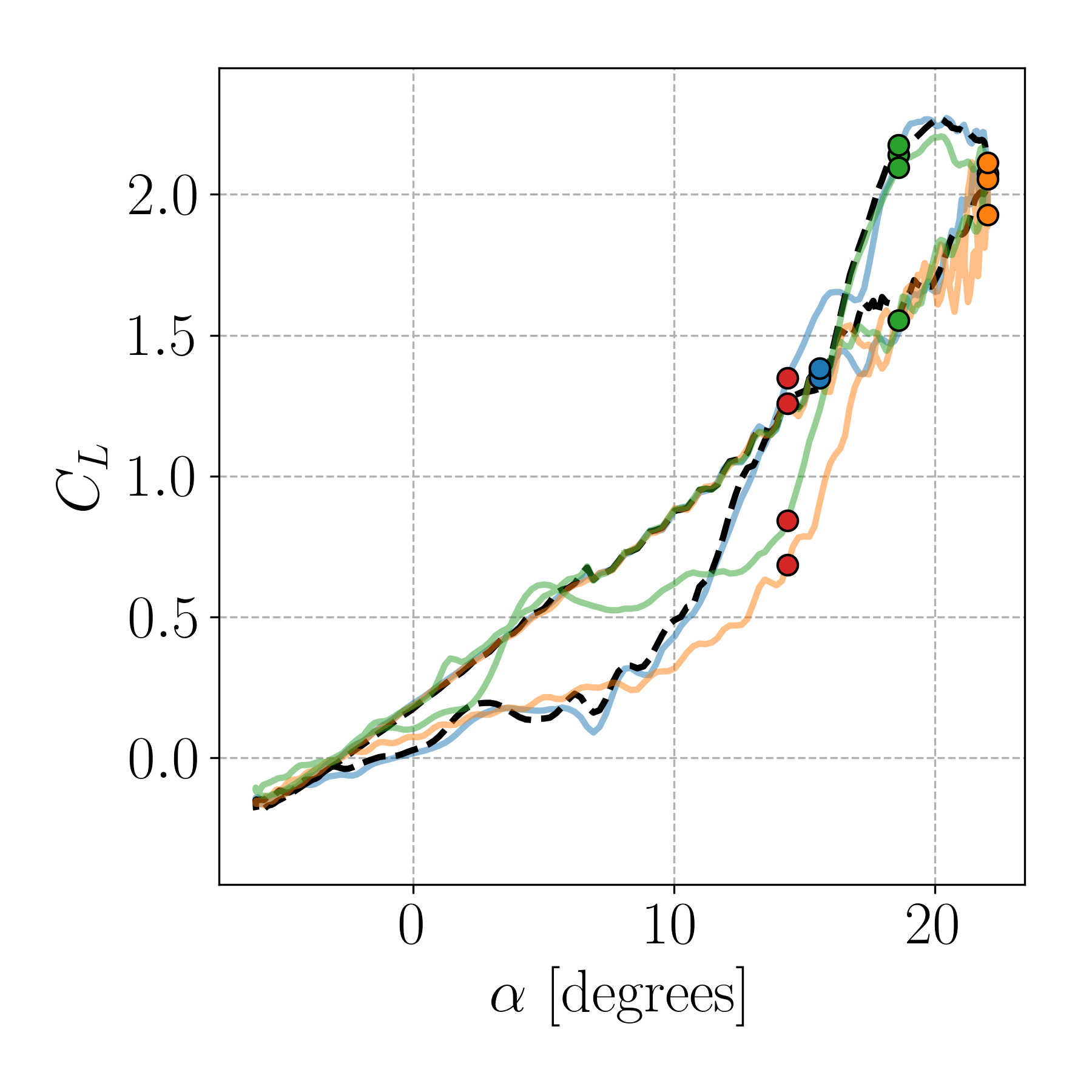}
		\end{subfigure}
		%
		\begin{subfigure}[b]{0.32 \textwidth}
			\centering
			\includegraphics[width=1\textwidth]{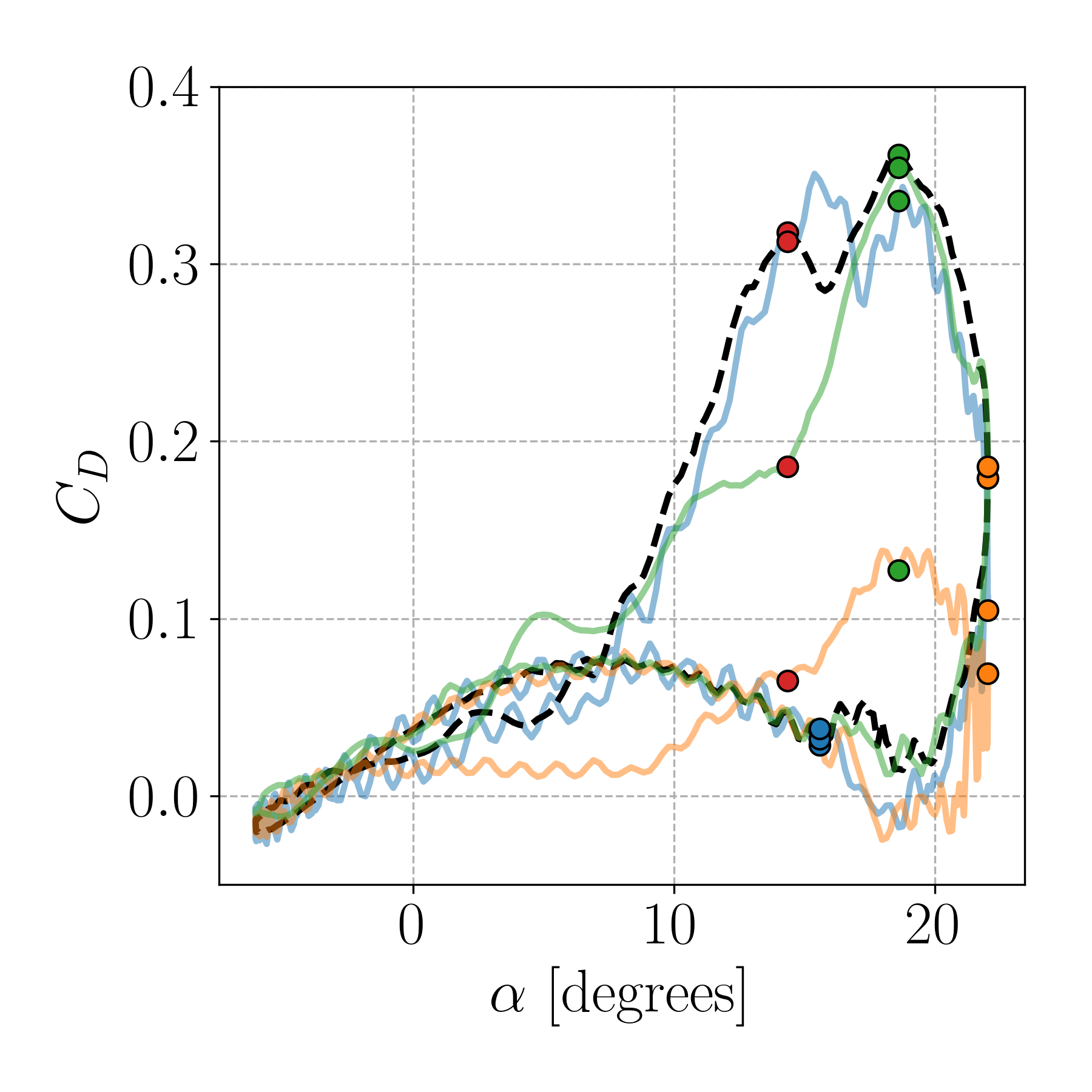}
		\end{subfigure}
		%
		\begin{subfigure}[b]{0.32 \textwidth}
			\centering
			\includegraphics[width=1\textwidth]{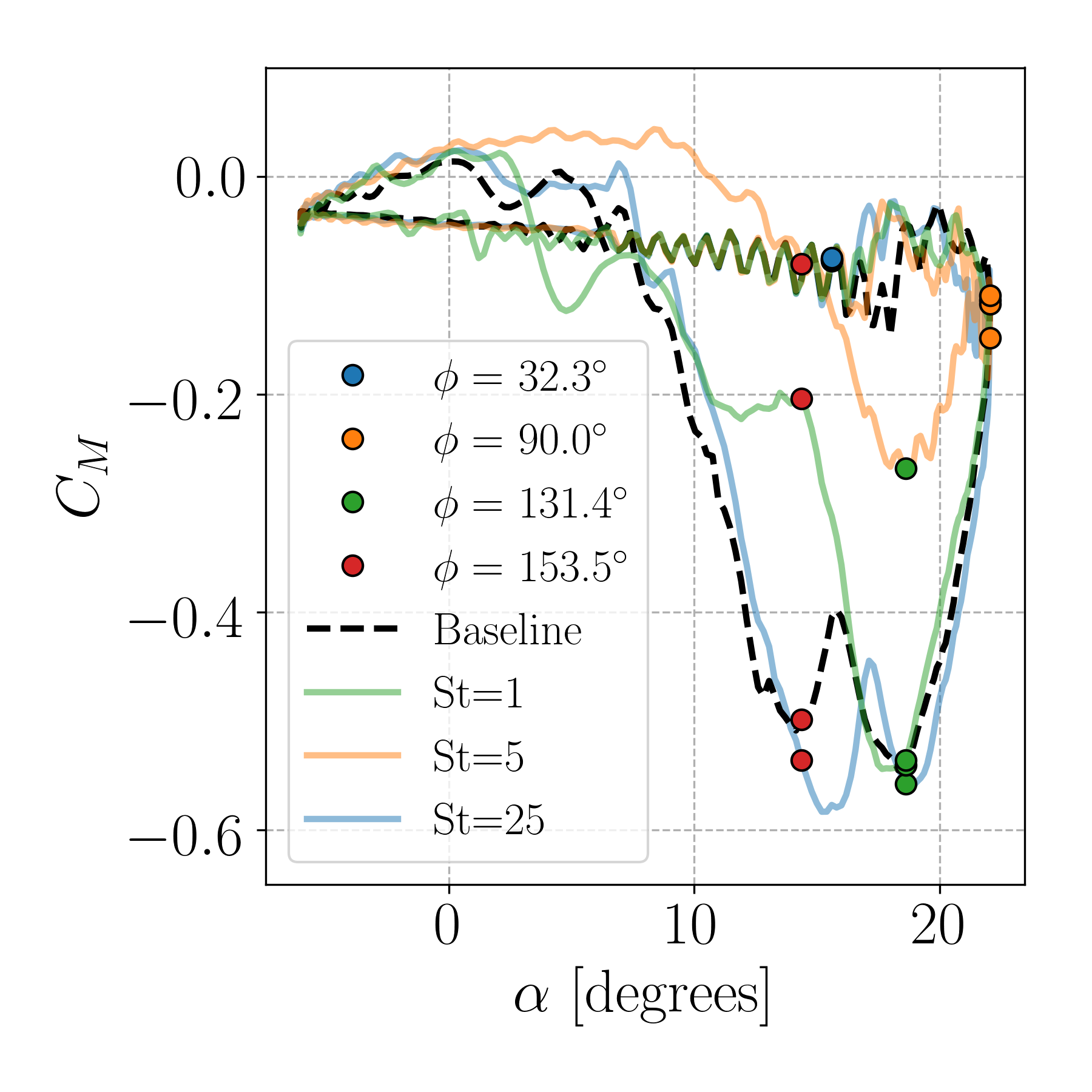}
		\end{subfigure}		
		\caption{Aerodynamic coefficients versus effective angle of attack for 2D actuated flows with different frequencies (Case 2).}	
		\label{fig:comparison_control_cases2}	
	\end{figure*}

	In what follows, results will be discussed based only on ``Case 2'' flow actuation. Figure \ref{fig:comparison_control_cases2} shows plots of aerodynamic coefficients as functions of the effective angle of attack. Results of the baseline flow are compared to those with actuation for $ St = 1$, $5$ and $25$. Hence, it is possible to evaluate the effects of low, moderate and high frequencies of actuation on the aerodynamic loads during any instant of the motion. It is clear that the actuation frequency has a large impact in the flow response, especially for instants of downward velocity.
	
	Different moments of the plunging motion are also highlighted by circles at $ \phi = 32.3^{\circ}$, $90.0^{\circ}$, $131.4^{\circ}$ and $153.5^{\circ}$. One should be reminded from Fig. \ref{fig:phase_angles} that $\phi \in [0^\circ, 180^\circ]$ represents the downstroke motion which includes the formation, transport and ejection of the leading-edge vortex.
	These specific values of $ \phi$ are shown due to important flow features that occur at such instants and that will be used to compare the actuation setups next.
	
\begin{figure*}
			\centering
			\includegraphics[width=\textwidth]{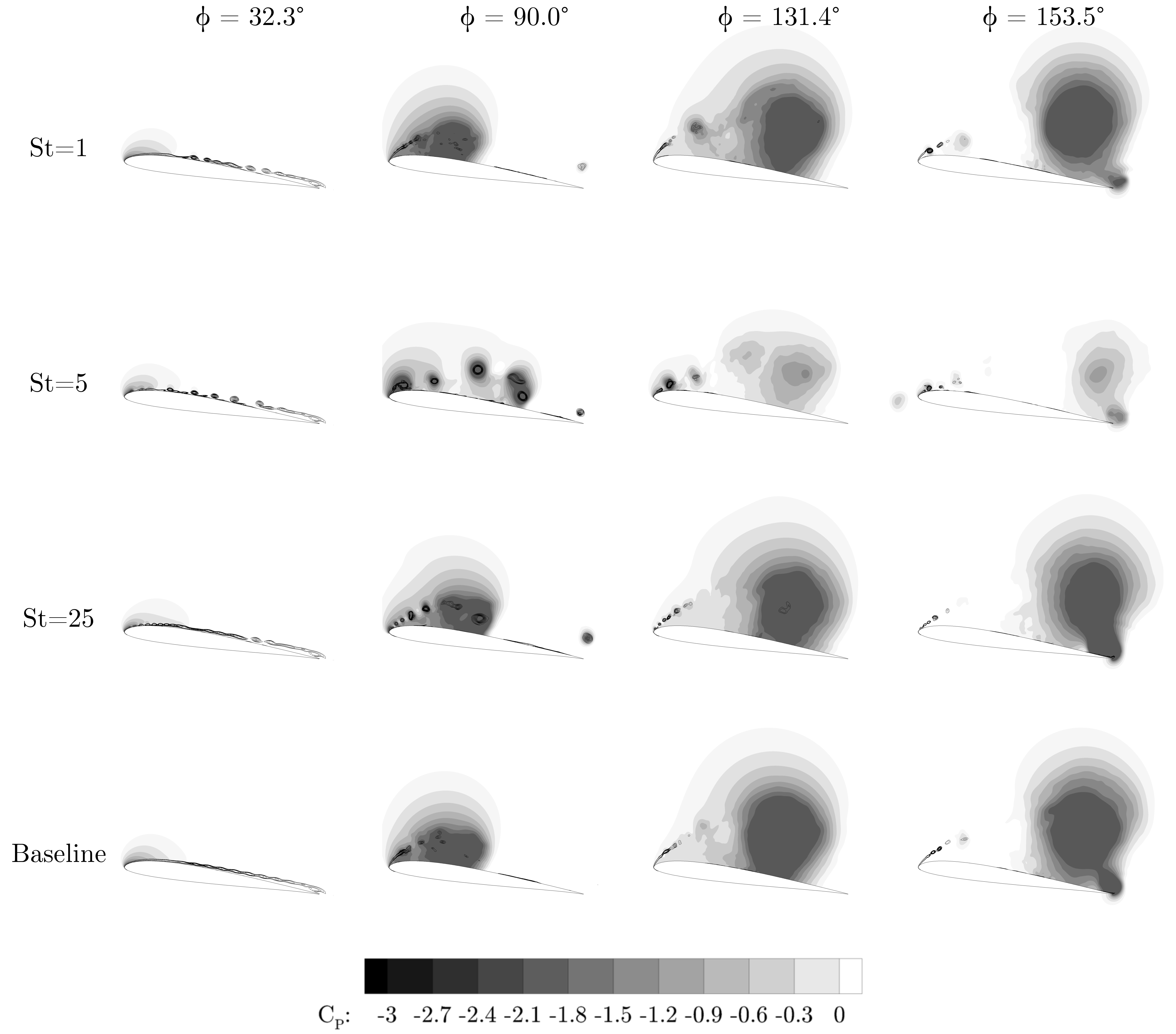}
			\caption{ $ C_P $ contours with iso-lines of z-vorticity for spanwise-averaged flows with 2D actuation (Case 2).}
			\label{fig:cp_comparison1}	
		\end{figure*}

	Contours of spanwise-averaged pressure coefficient $ C_P $ with iso-contours of z-vorticity are shown in Fig. \ref{fig:cp_comparison1} and Supplemental Material \cite{sup2} for the same actuation frequencies as in Fig. \ref{fig:comparison_control_cases2} and for the baseline case. It is observed that actuation does not delay the formation of the dynamic stall vortex but disrupts it. At $ \phi = 32.3^{\circ}$, all flows have roughly the same aerodynamic loads (notice that the blue circles lie on top of each other in Fig. \ref{fig:comparison_control_cases2}). However, the shear layer is clearly disrupted by actuation, especially in the $ St = 5 $ case. When compared to the baseline case, it can be seen that Kelvin-Helmholtz instabilities appear and grow earlier in the plunging motion for the $ St = 5 $ setup. Actuation at other frequencies also modify the shear-layer but instabilities do not get amplified as much. At $ \phi = 90.0^{\circ}$ (maximum downward velocity), the formation of the leading-edge vortex does not occur as prominently in the $ St = 5 $ case when compared to other actuation frequencies. For this case, vortices created by actuation successfully break the large-scale coherent structure formed at the leading edge. On the other hand, for $ St = 1 $ and  $ 25 $, the vortices created by the actuation do not effectively disrupt the formation of the LEV. In the latter case, small vortical structures end up coalescing and forming the LEV in a similar fashion compared to the baseline flow. 
		\def \sizesubfigCp{0.16}

	At $131.4^{\circ}$ we observe in Fig. \ref{fig:comparison_control_cases2} the highest value of $ C_D $ for the baseline flow. Actuated flows exhibit similar aerodynamic coefficients, except for $ St = 5 $. At this frequency, Fig. \ref{fig:cp_comparison1} shows a coherent structure with higher (less negative) values of $ C_P $ compared to other cases. This effect is a consequence of the formation of smaller vortical structures by the actuation that do not coalesce into a single dynamic stall vortex at first. This weaker LEV also induces the formation of a less intense TEV at $ \phi = 153.5^{\circ}$. This latter instant is represented in Fig. \ref{fig:comparison_control_cases2} by a second peak in drag coefficient for the baseline flow.
		
\begin{figure}[ht]
		\centering
		\begin{subfigure}{0.49\textwidth}
			\includegraphics[width=\textwidth]{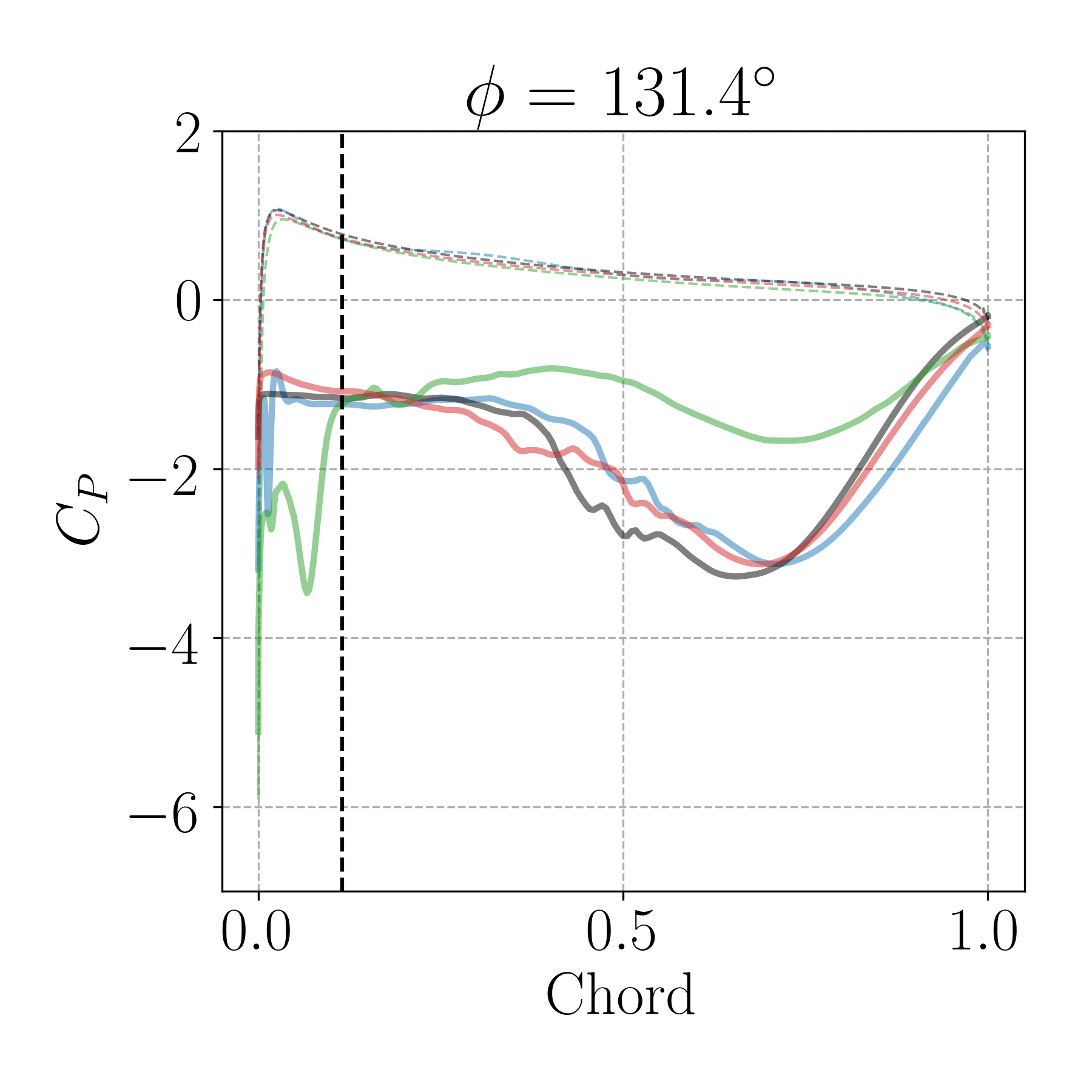}
		\end{subfigure}
		\begin{subfigure}{0.49\textwidth}
			\includegraphics[width=\textwidth]{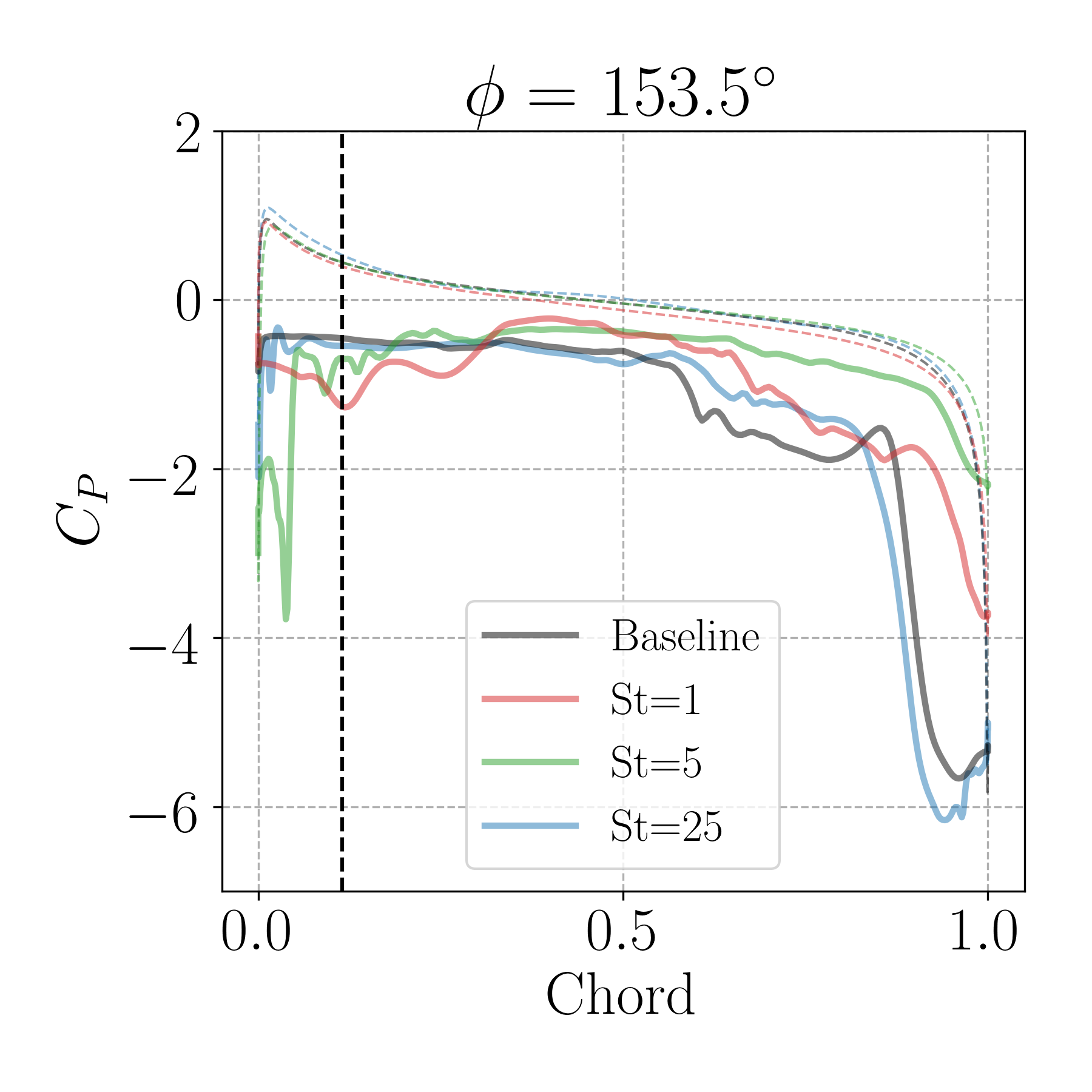}
		\end{subfigure}		
		\caption{Comparison between span-averaged values of $ C_P $ for 2D actuators with different frequencies (Case 2). The vertical dashed line indicates the location of $ x_{vsn} $.}
		\label{fig:cpCurves}
	\end{figure}

\begin{figure}[ht]
			\centering
			\includegraphics[width=0.6\textwidth, trim={0 0 0 300px}, clip]{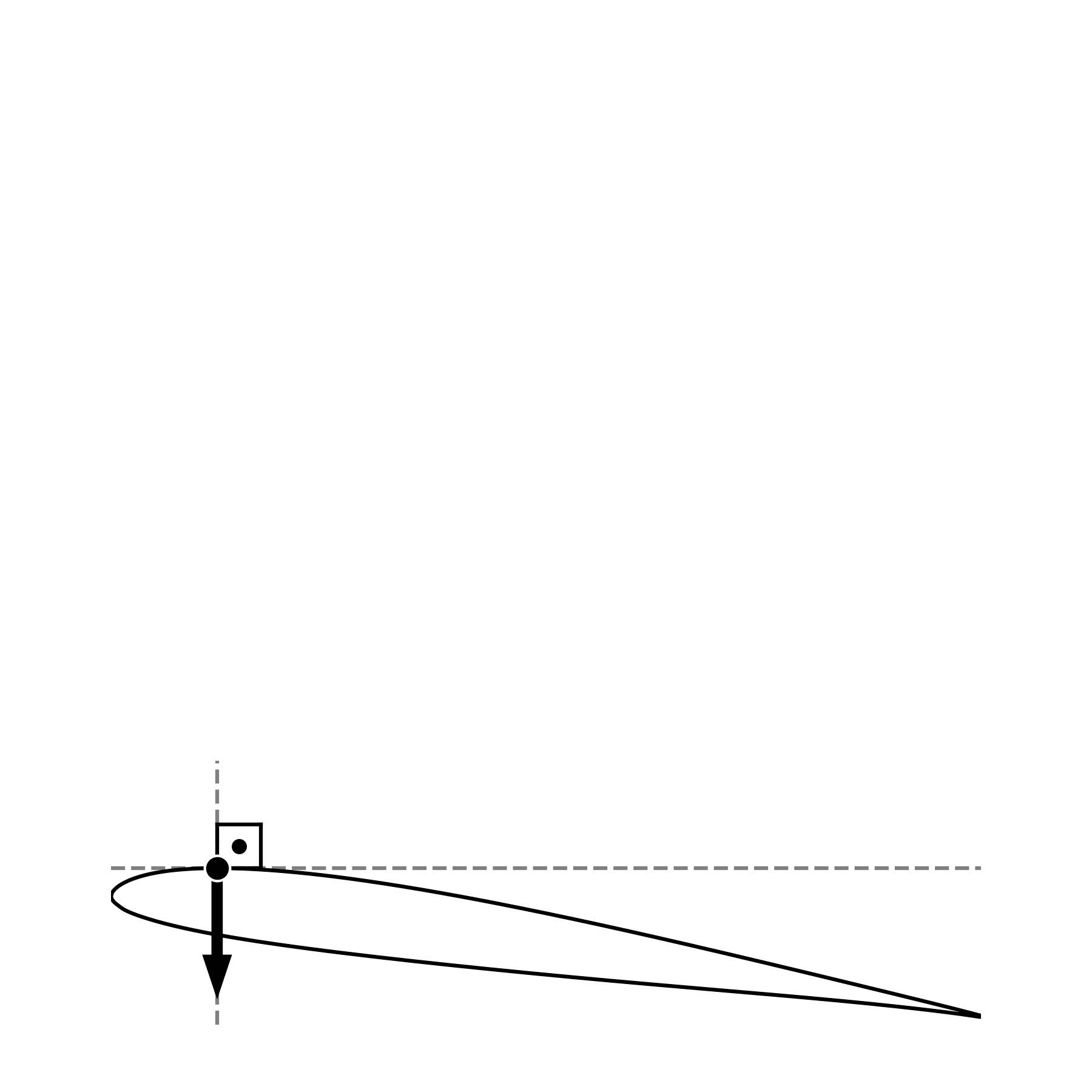}
			\caption{Position $ x_{vsn}$ where the inward pointing surface normal at the suction side is vertical.}
			\label{fig:xvn}
		\end{figure}

	Figure \ref{fig:cpCurves} shows $ C_P $ distributions (spanwise-averaged) in order to better quantify pressure differences among the various flows previously analyzed. Results are presented at $ \phi = 131.4^{\circ}$ and $153.5^{\circ}$ as a function of the airfoil chord location. A vertical dashed line marks the position where the surface normal on the airfoil wall (on the suction side) is vertical, as shown in Fig. \ref{fig:xvn}. This position is given by $ x_{vsn} = 0.12 L $ and it is important to differentiate how the regions over the airfoil suction side contribute to drag reduction.
	We consider the surface normal pointing inward the airfoil. Lift and drag generated from pressure distributions along the airfoil surface are calculated by $L=\oint p \,n_y\, dS$ and $D=\oint p\, n_x\, dS$, respectively. Here $n_x$ is the component of surface normal in the $x$ direction while $n_y$ is that in the $y$ direction. Thus, a force applied in the normal direction on the airfoil suction side, to the left of the vertical dashed line,  lead to lift reduction and drag increase. On the other hand, a normal force applied to the right of such line result in both lift and drag reductions. Pressure forces applied on the bottom side of the airfoil will always lead to lift increase.
		
	For the baseline case, at $ \phi = 131.4^{\circ}$, the bump in the $ C_P $ distribution appears due to the advection of the LEV over the suction side of the airfoil. This negative value of pressure coefficient indicates that a strong suction occurs on the top surface of the airfoil, leading to a lift and drag increase. Similar trends are observed for the cases with $St=1$ and $25$. At $ \phi = 153.5^{\circ}$, a strong suction peak is observed at the trailing edge due to the formation of the TEV and such feature also increases both lift and drag. Again, the solution obtained for $St=25$ is very similar to that from the baseline flow. On the other hand, for the $St=5$ setup, one observes that a mild bump forms at $ \phi = 131.4^{\circ}$, reducing both lift and drag for this case. However, a strong suction peak is present at the leading edge of the airfoil, increasing both lift and further reducing drag. When the airfoil is at $ \phi = 153.5^{\circ}$, a suction peak is still present at the airfoil leading edge and a minor suction effect is observed at the trailing edge due to a less intense TEV. In summary, lower (more negative) values of $ C_P $ to the left of the vertical dashed line in Fig. \ref{fig:cpCurves} would result in lower pressure drag. In the same context, higher (less negative) values of $ C_P $ to the right of the vertical dashed line also lead to lower pressure drag. Both conditions are met when flow actuation is applied at $St=5$.
	
\begin{figure}[ht]
		\centering
		\begin{subfigure}{0.49\textwidth}
			\centering
			\includegraphics[width=\textwidth,trim={25px 25px 25px 25px },clip]{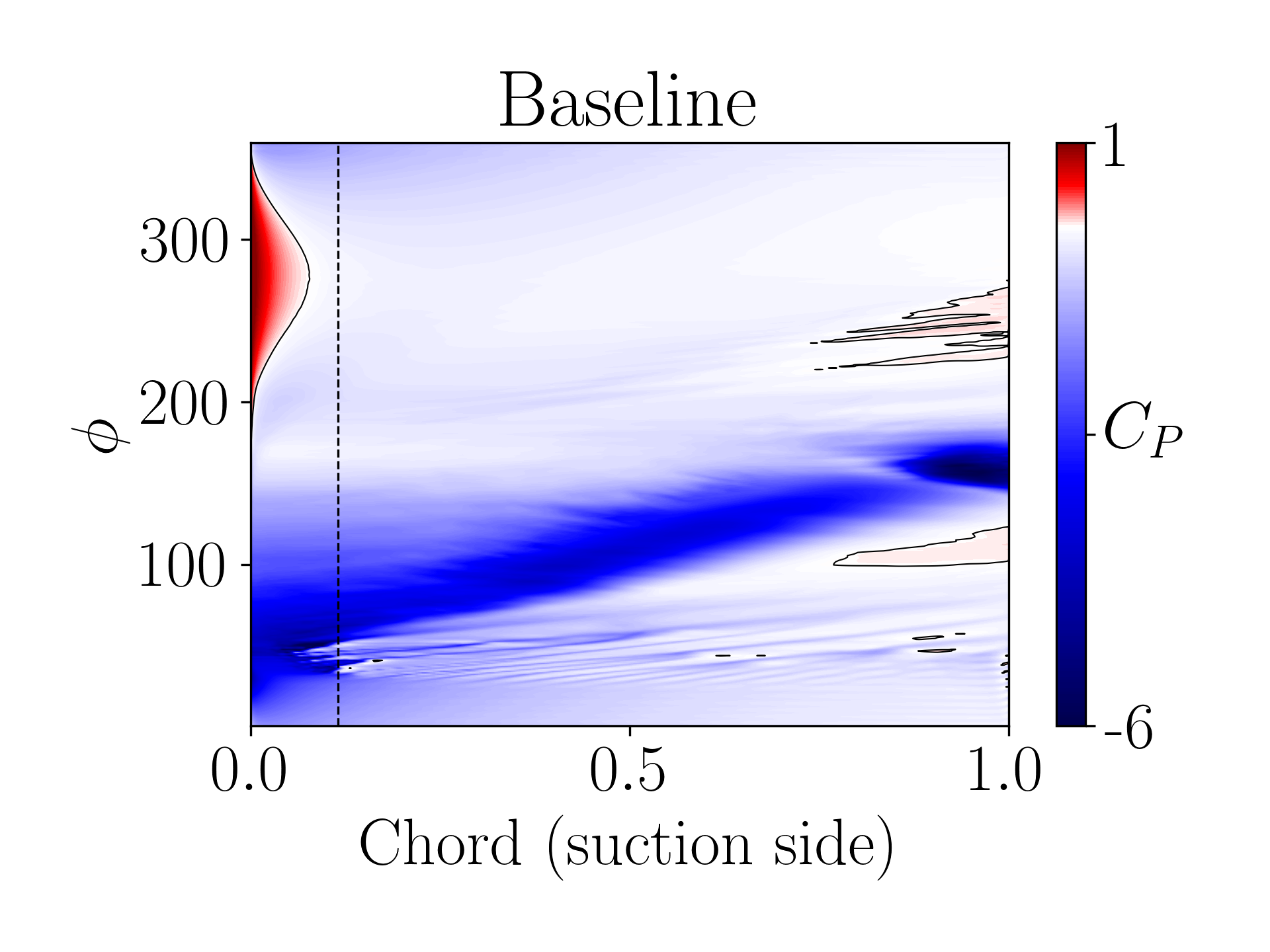}
		\end{subfigure}
		\begin{subfigure}{0.49\textwidth}
			\centering
			\includegraphics[width=\textwidth,trim={25px 25px 25px 25px },clip]{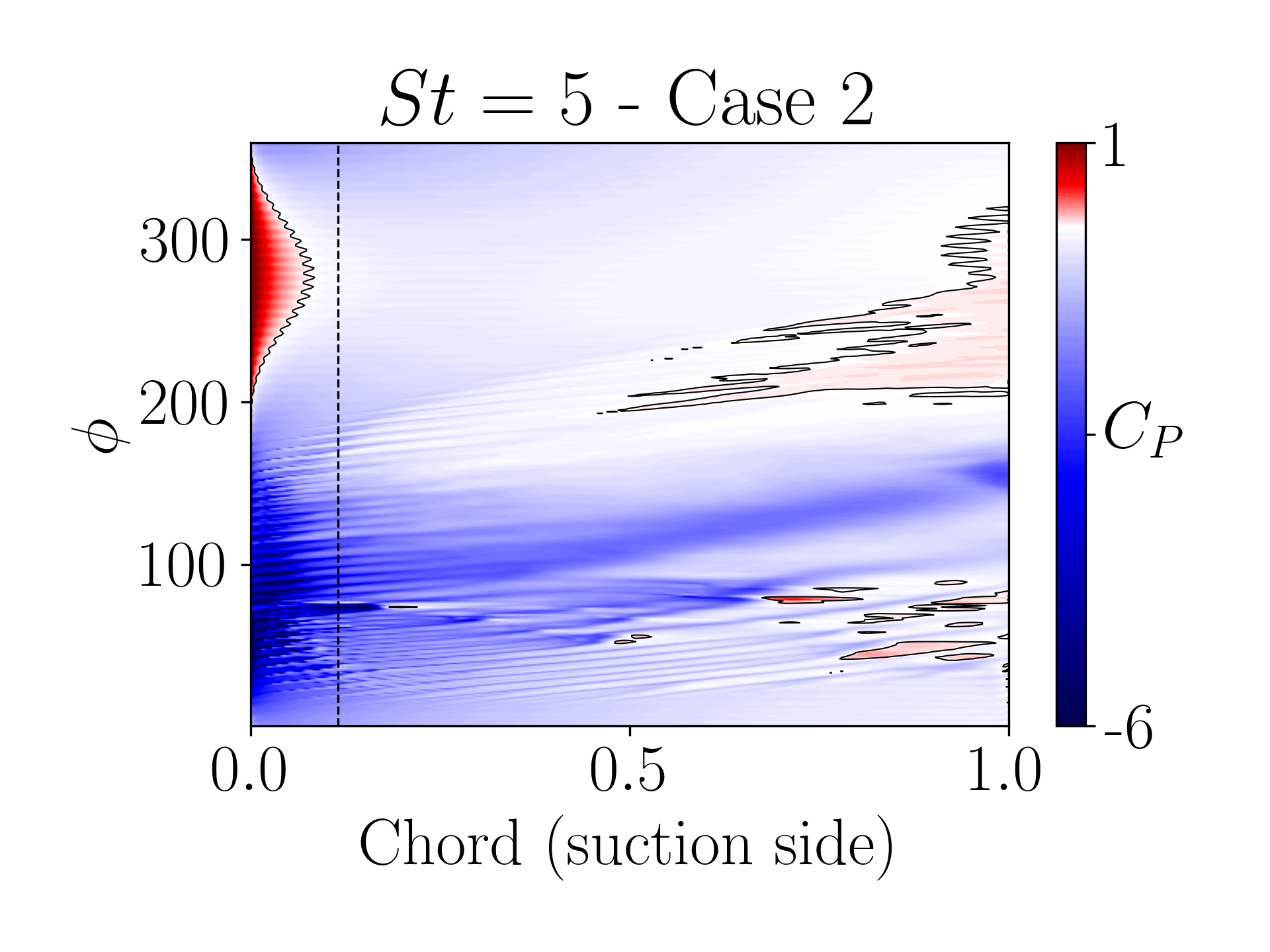}
		\end{subfigure}
		\caption{Comparison of $ C_P $ between baseline and 2D actuated flow with $ St=5 $.}
		\label{fig:cpmap}
	\end{figure}

\begin{figure}[ht]
		\centering
		\begin{subfigure}{0.49\textwidth}
			\centering
			\includegraphics[width=\textwidth,trim={25px 25px 25px 25px },clip]{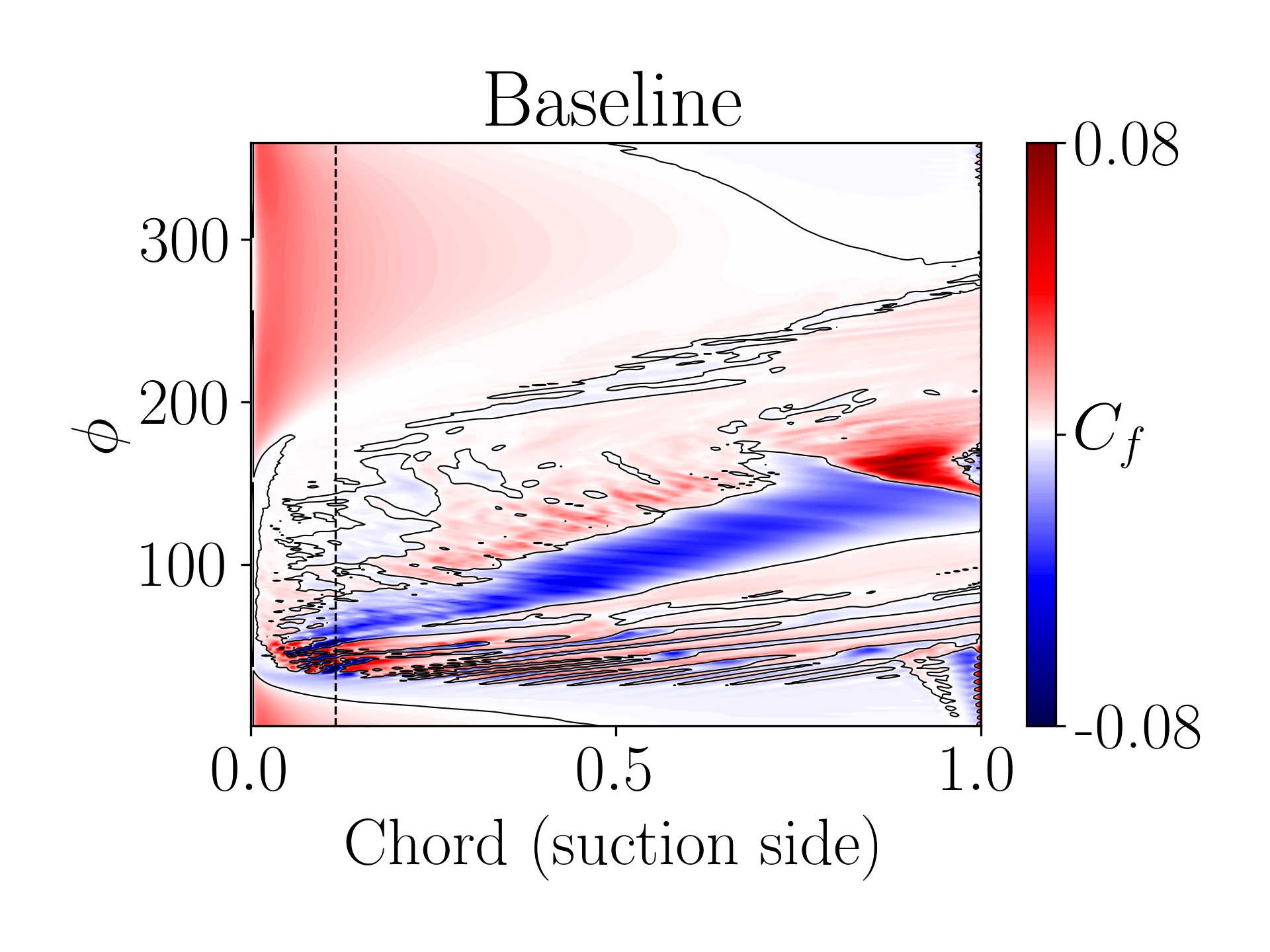}
		\end{subfigure}
		\begin{subfigure}{0.49\textwidth}
			\centering
			\includegraphics[width=\textwidth,trim={25px 25px 25px 25px },clip]{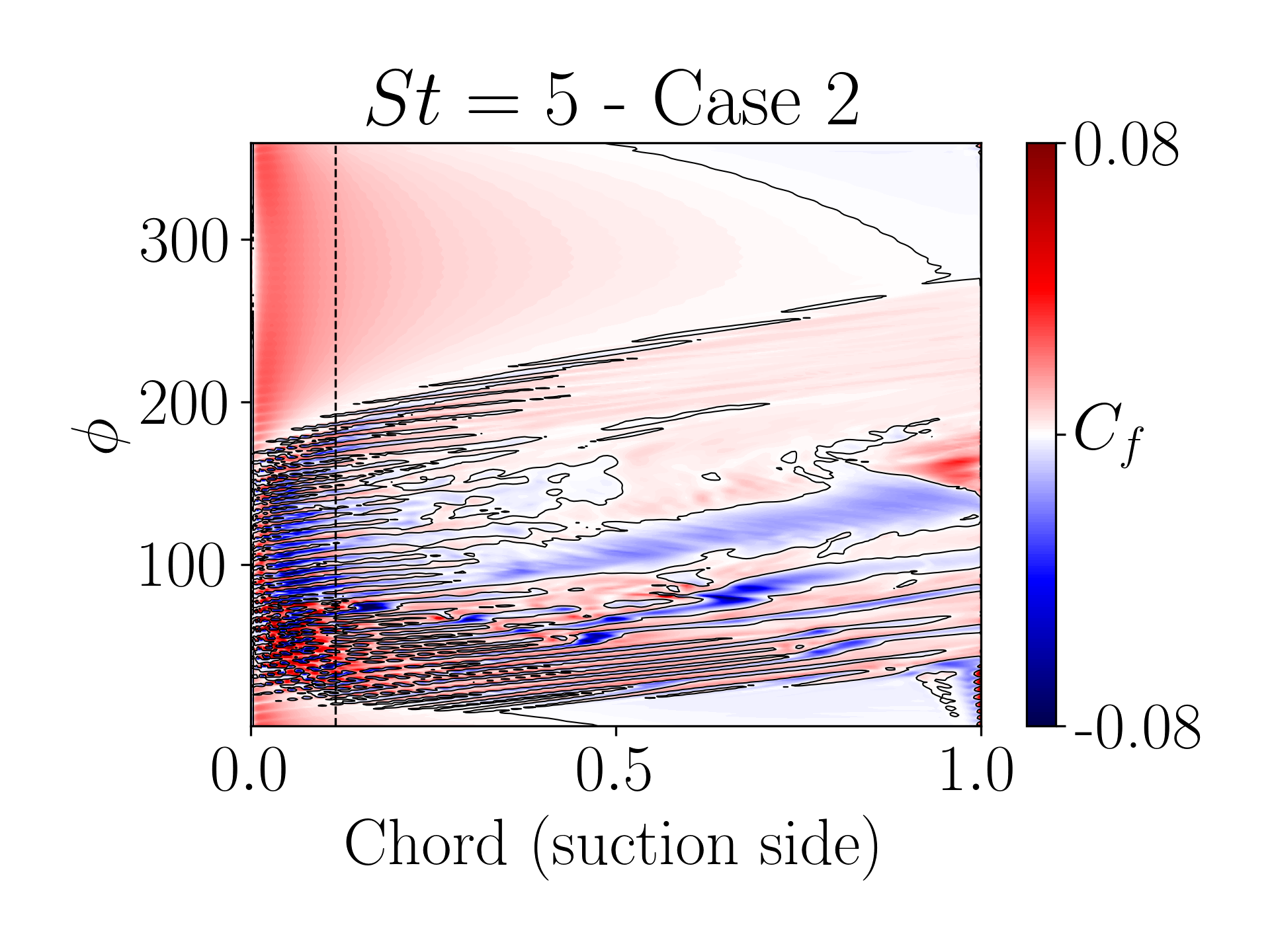}
		\end{subfigure}
		\caption{Comparison of $ C_f $ between baseline and 2D actuated flow with $ St=5 $.}
		\label{fig:cfmap}
	\end{figure}

	The full history of spanwise-averaged $ C_P $ computed on the airfoil suction side is displayed in Fig. \ref{fig:cpmap} as a function of $ \phi $. In this figure, a comparison is shown for the baseline and $St = 5$ cases. The dark blue colors in the plots represent the low pressure signatures from the LEV and TEV and one can see that they are less severe in the case with control. Figure \ref{fig:cfmap} shows similar maps but colored by friction coefficient $ C_f $ instead. For lower $ \phi $ angles, it is possible to notice the oscillatory behavior of $C_f$ due to the initial shear layer instabilities. The dark blue contours mark the separation region caused by the transport of the LEV while the dark red contours in the trailing edge are due to formation of the TEV. In the case with actuation, the LEV is weaker so the blue trace is thinner and less intense than that computed for the baseline configuration. From this figure, it is also possible to see that the separation near the leading edge has an oscillatory behavior due to flow actuation during the downstroke motion.
	\begin{figure}[ht]
		\centering
		
		\begin{subfigure}[t]{\3PerLine \textwidth}
			\centering
			\includegraphics[width=\textwidth]{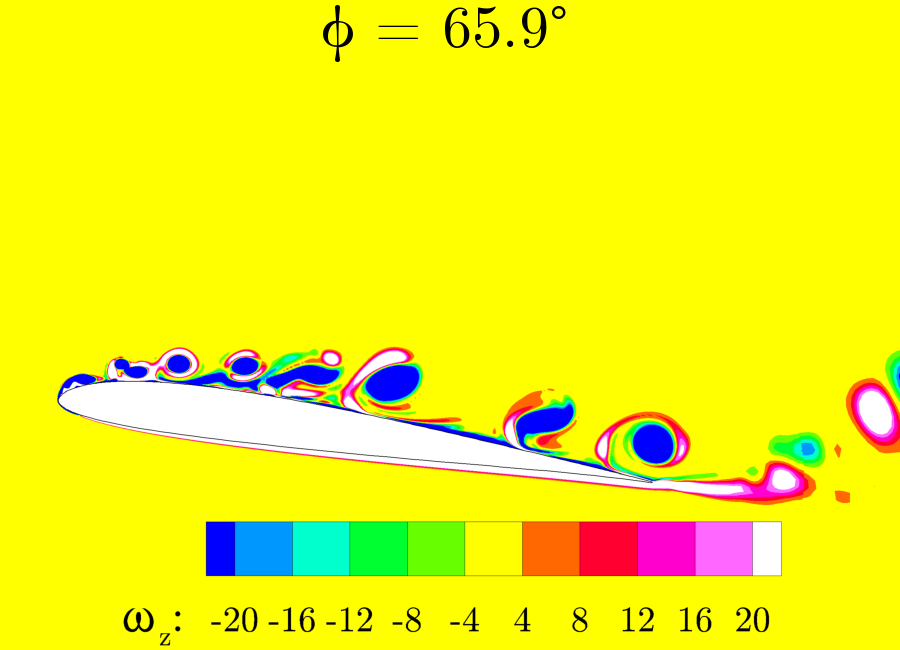}
		\end{subfigure}	
		\begin{subfigure}[t]{\3PerLine \textwidth}
			\centering
			\includegraphics[width=\textwidth]{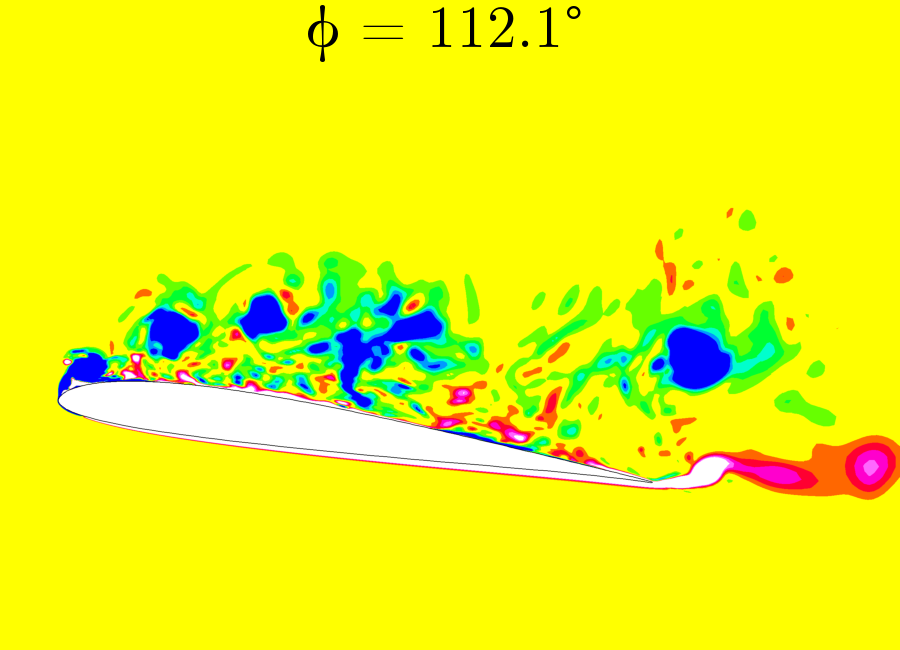}
		\end{subfigure}	
		\begin{subfigure}[t]{\3PerLine \textwidth}
			\centering
			\includegraphics[width=\textwidth]{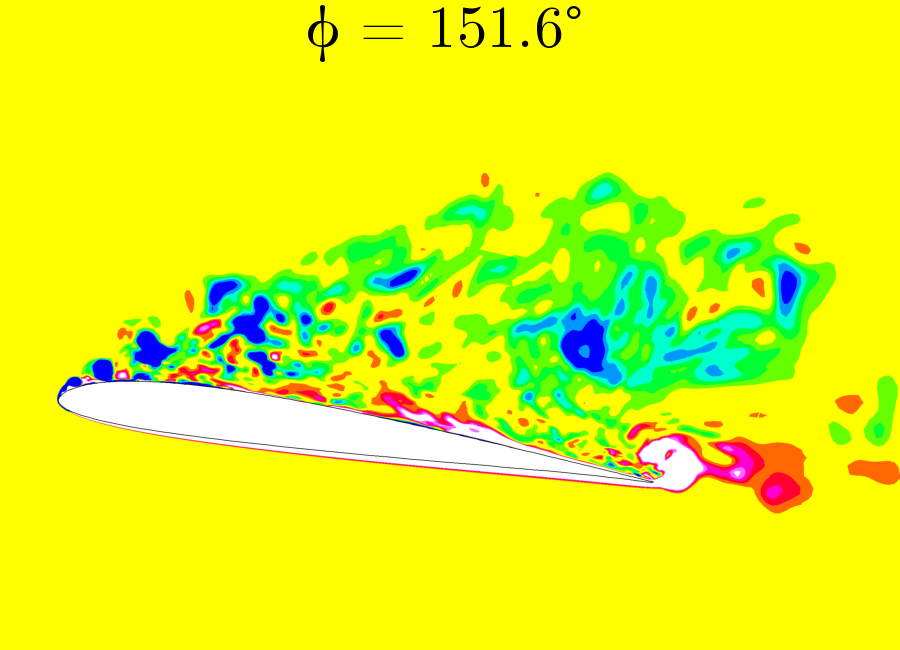}
		\end{subfigure}		
		\caption{Spanwise-averaged vorticity contours at different phase angles for the $St$ = 5 controlled case (Case 2).}
		\label{fig:z_vorticity_control}
	\end{figure}
	
	The $St = 5$ actuation leads to a disruption of the LEV which sheds small pockets of vorticity instead of accumulating it. This effect can be observed in Fig. \ref{fig:z_vorticity_control} and Supplemental Material \cite{sup1} and it avoids the formation of a large-scale coherent structure at the leading edge, in contrast to the baseline configuration. 	
	In summary, a significant reduction in $ C_D $ and $ C_M $ occurs as a result of the features observed due to flow actuation: mitigation of the dynamic stall vortex, strong negative values of $ C_P $ upstream of $ x_{vsn} $, and mild values of $ C_P $ downstream of $ x_{vsn} $ for $ 2.5 < St < 15 $. Although the flow actuation leads to a small reduction in terms of $ C_L $, it is not as prominent as the reductions observed in $ C_D $ and $ C_M $. Since a lower actuation disturbance is employed for Case 2, and the best result in terms of $ \frac{\overline{C_L}}{\overline{C_D}} $ for this case is obtained for $St=5$, we will further investigate this specific flow configuration. Therefore, we can reduce the energy expenditure in the actuation while maintaining the mean lift to mean drag ratio above 20.
	
	\subsection{3D Actuation}
	
	In the previous section, results of 2D flow actuation for the present plunging airfoil were presented. However, results shown in the literature discuss the enhanced performance of 3D actuation for drag reduction in airfoil flows involving static stall \cite{Chi-An:2019, Munday:2018}. Therefore, we present a study of different configurations of 3D actuation to assess their impact on drag reduction. Results are shown for $ St = 5 $ and $ C_{\mu} $ from Case 2 for the actuation configurations discussed in Section \ref{control}.
		\begin{figure}[ht]
			\centering
			\includegraphics[width=\textwidth]{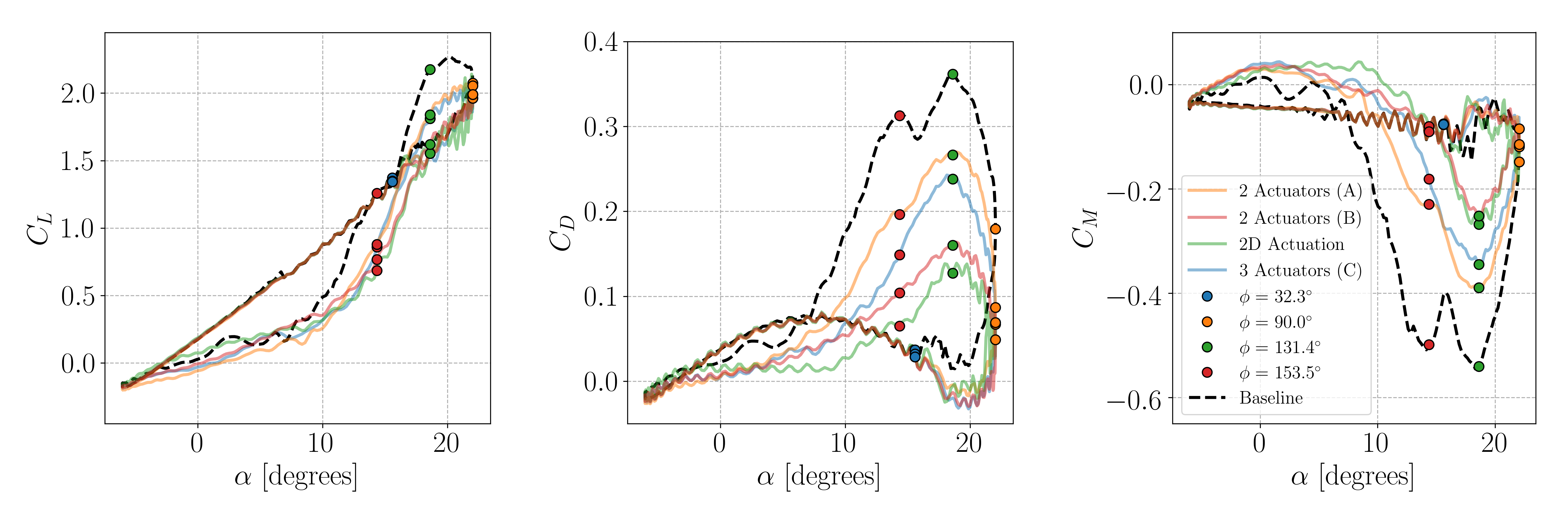}	
			\caption{Comparison of aerodynamic coefficients obtained by 2D and 3D actuation with $ St = 5 $ (Case 2).}	
			\label{fig:cxs3D}	
		\end{figure}
		
	In Fig. \ref{fig:cxs3D}, results are shown for the aerodynamic coefficients and it can be seen that all cases with 3D actuation exhibit higher values of $ C_L $ for high effective angles of attack $\alpha$ when compared to the 2D actuated flow. However, the values of $ C_D $ are considerably lower for the 2D actuation at the same angles of attack. The same can be said for $ C_M $, except for the case with two larger slots (configuration B), which has comparable values of moment coefficient to those obtained for the 2D actuation.
	\begin{figure}[ht]
		\centering
		\includegraphics[width=\textwidth]{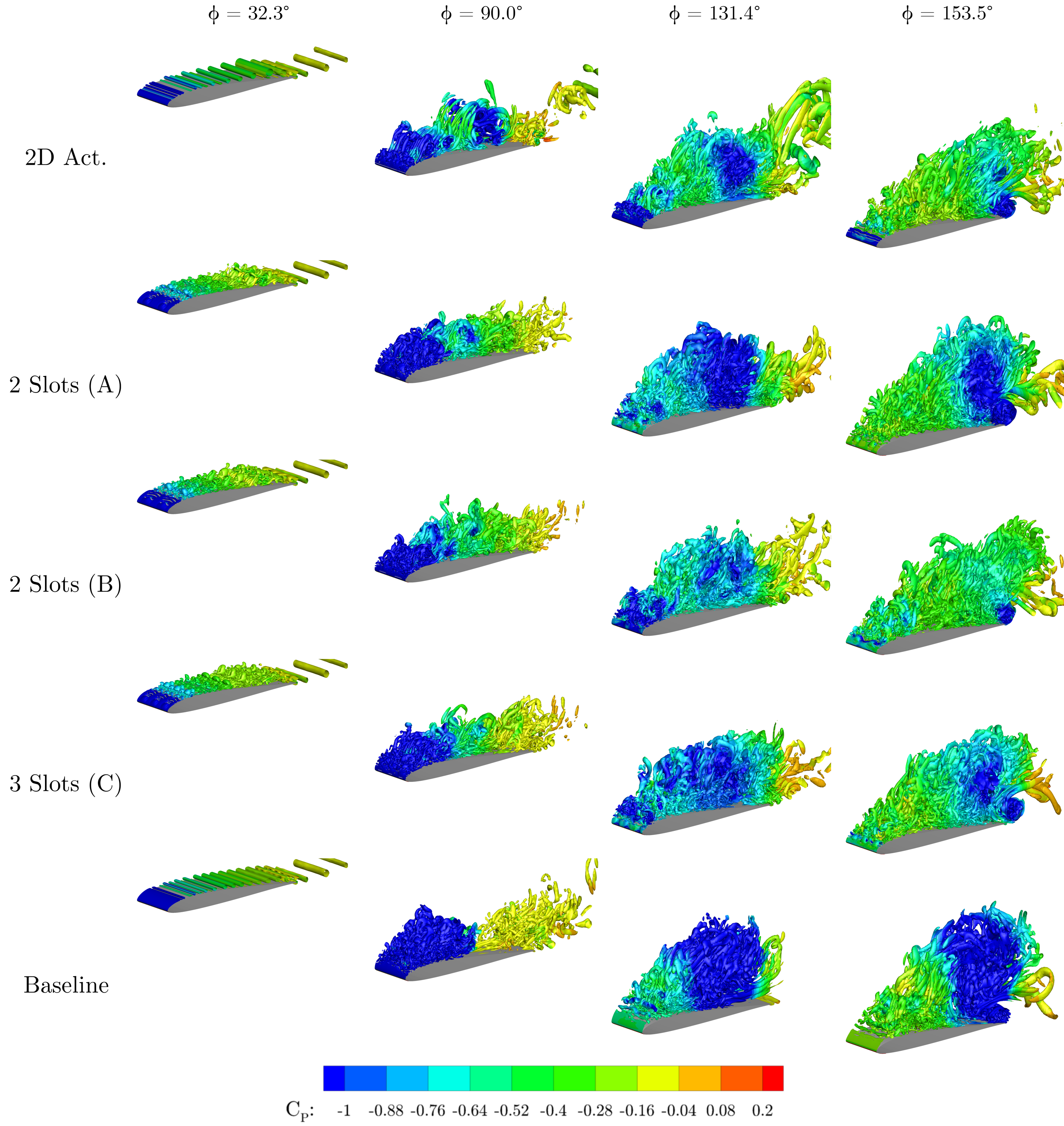}
		\caption{Q-criterion colored by $ C_P $ comparing 2D and 3D actuation with $ St = 5 $ (Case 2) at various phases of the plunge motion.}
		\label{fig:qcriterion3DAct}
	\end{figure}	
	
	Iso-surfaces of Q-criterion colored by pressure coefficient are shown in Fig. \ref{fig:qcriterion3DAct} at various moments of the plunge motion. A movie with the same features is presented as Supplemental Material \cite{sup3} comparing 2D and 3D actuations. Due to its inherent three-dimensionality, 3D actuation exhibits earlier transitional features at $ \phi = 32.3^{\circ}$ when compared to the baseline and 2D actuation cases. All actuated flows exhibit weaker LEVs compared to the baseline, noting that 2D actuation is the most efficient since it is able to efficiently disrupt the LEV formation at $ \phi = 90.0^{\circ} $. At $\phi = 131.4^{\circ}$, we can notice that both  2D actuation and that with two wider slots (B) produce dynamic stall vortices with higher values (less negative) of pressure coefficient. With weaker LEVs, these cases also show TEVs which are less intense, avoiding the secondary drag peak that appears for the baseline configuration in Fig. \ref{fig:cxs3D} at $ \phi = 153.5^{\circ}$.
	
\begin{figure}[ht]
		\centering
		\includegraphics[width=\textwidth]{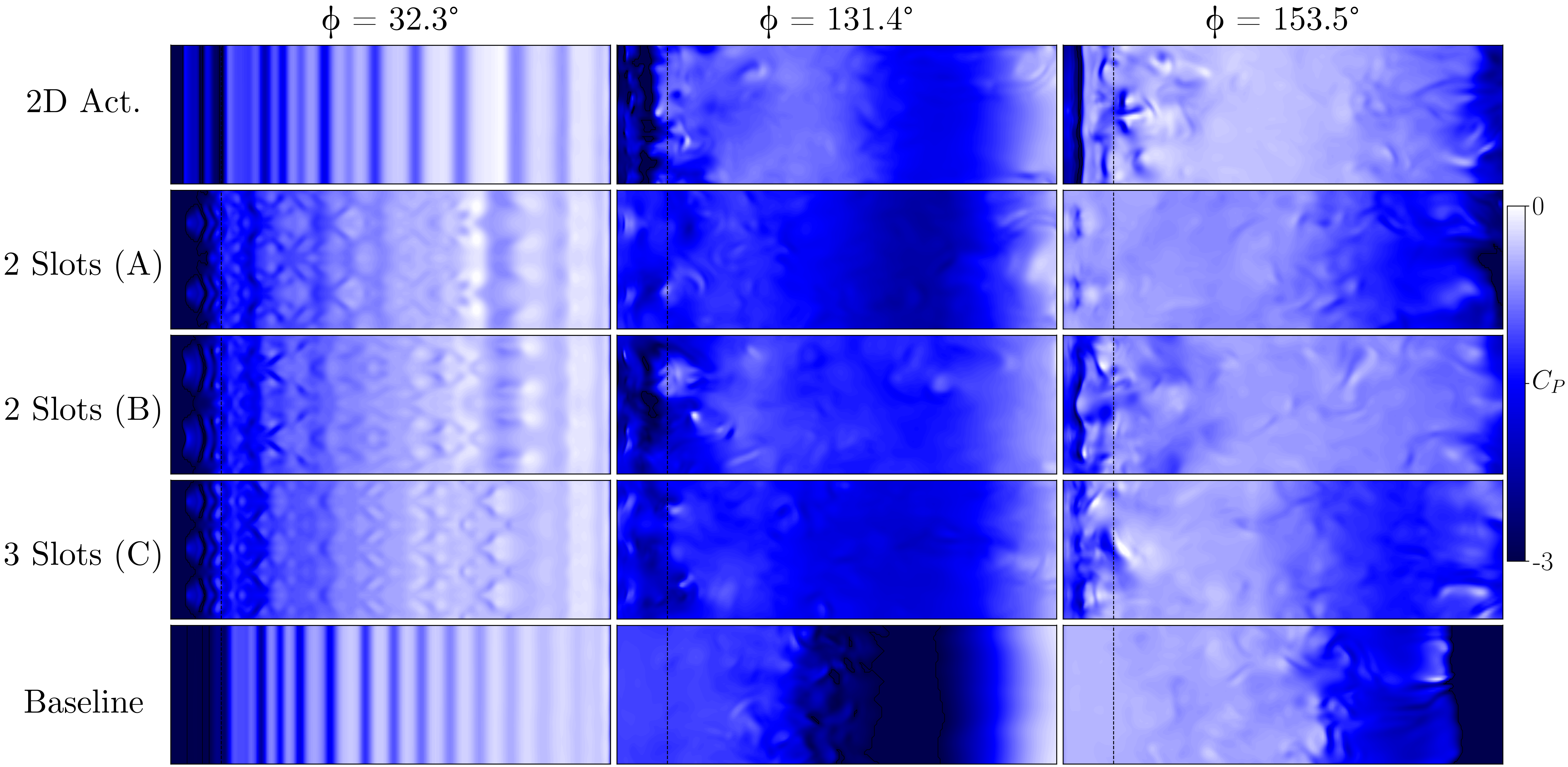}
		\caption{Distribution of $C_P$ over the airfoil suction side (flow is directed from left to right) for 2D and 3D actuation with $St = 5$ (Case 2). }
		\label{fig:cp3D}
	\end{figure}

	The impact of different types of actuation on $ C_P $ distribution along the airfoil suction side can be seen in Fig. \ref{fig:cp3D} and Supplemental Material \cite{sup4}. At $ \phi = 32.3^{\circ}$, despite similar values of aerodynamic loads observed in Fig. \ref{fig:cxs3D}, $ C_P $ contours are fairly distinct. Two-dimensional coherent structures are present in the baseline and 2D actuation cases, while all 3D actuated flows exhibit more complex 3D structures which promote transition to turbulence earlier in the plunging motion. When $ C_D $ reaches its peak at $ \phi = 131.4^{\circ}$, a dark region of low pressure created by the LEV is present in the baseline flow, while milder values of $C_P$ are observed in the actuated cases. In general, the 2D actuated flow has less negative values of $ C_P $ downstream of $ x_{vsn} $ when compared to the other cases and the $C_P$ values are more negative upstream $ x_{vsn} $. Similar observations can be made at $ \phi = 153.5^{\circ}$ regarding the TEV.
	
\begin{figure}[ht]
		\centering
		\includegraphics[width=\textwidth]{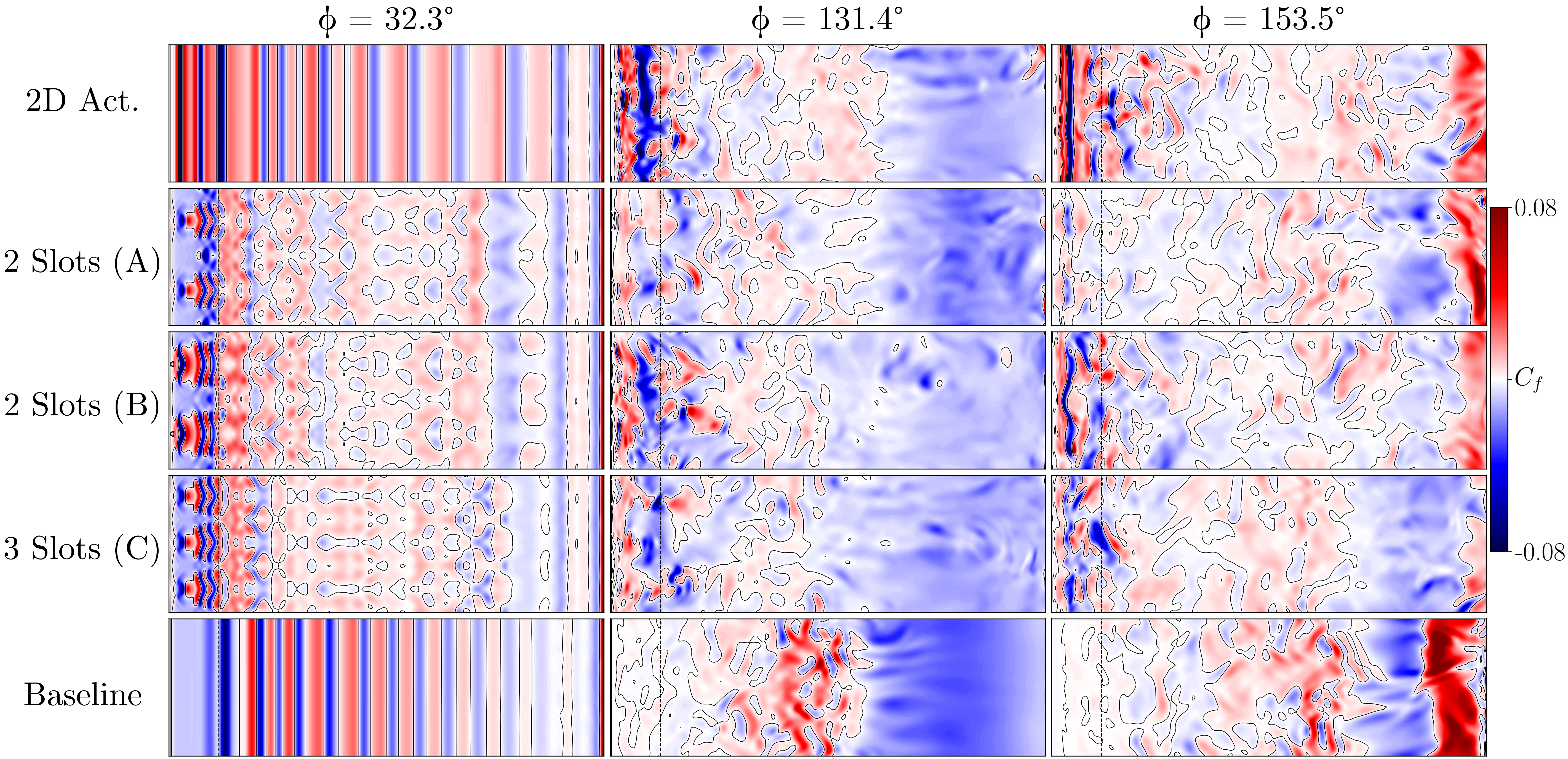}
		\caption{Distribution of $C_f$ over the airfoil suction side (flow is directed from left to right) for baseline and 2D and 3D actuation with $St = 5$ (Case 2).}
		\label{fig:cf3D}
	\end{figure}

	Figure \ref{fig:cf3D} and Supplemental Material \cite{sup5} show how flow separation changes due to actuation. While the flow is fully two-dimensional in the baseline and 2D actuated cases at $ \phi = 32.3^{\circ}$, the same cannot be said for the cases with 3D actuation. After transition takes place, regions of separation and reattachment upstream of $ x_{vsn} $ show higher spanwise coherence in the 2D actuated flow. Nevertheless, as can be seen at $ \phi = 131.4^{\circ}$, the separation created by the LEV is attenuated in all control cases. At $ \phi = 153.5^{\circ} $, all the actuated flows are able to form the TEV further downstream compared to the baseline case, reducing its overall impact on the aerodynamic coefficients.
	
\begin{figure}[ht]
		\centering
		\begin{subfigure}{0.49\textwidth}
			\includegraphics[width=\textwidth]{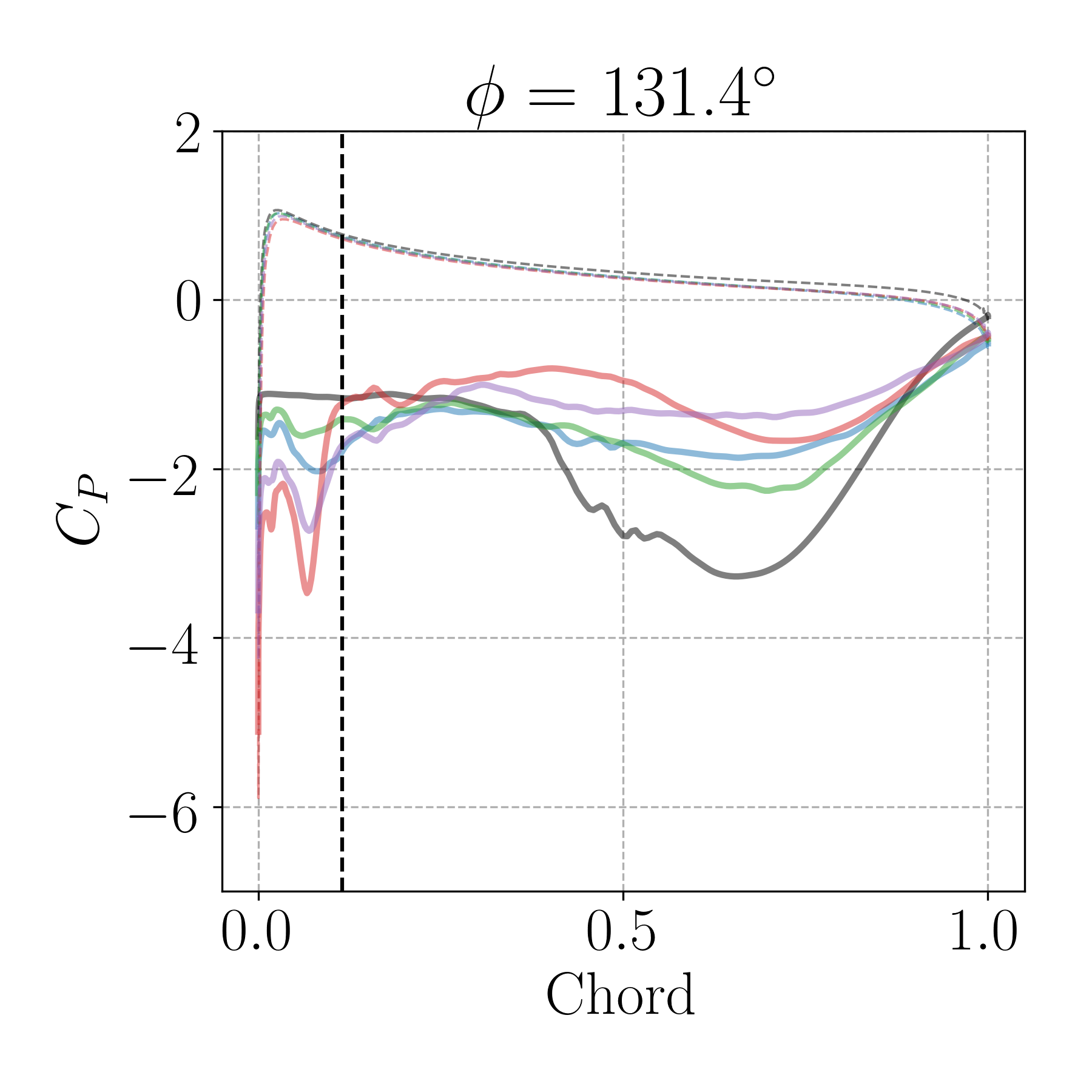}
		\end{subfigure}		
		\begin{subfigure}{0.49\textwidth}
			\includegraphics[width=\textwidth]{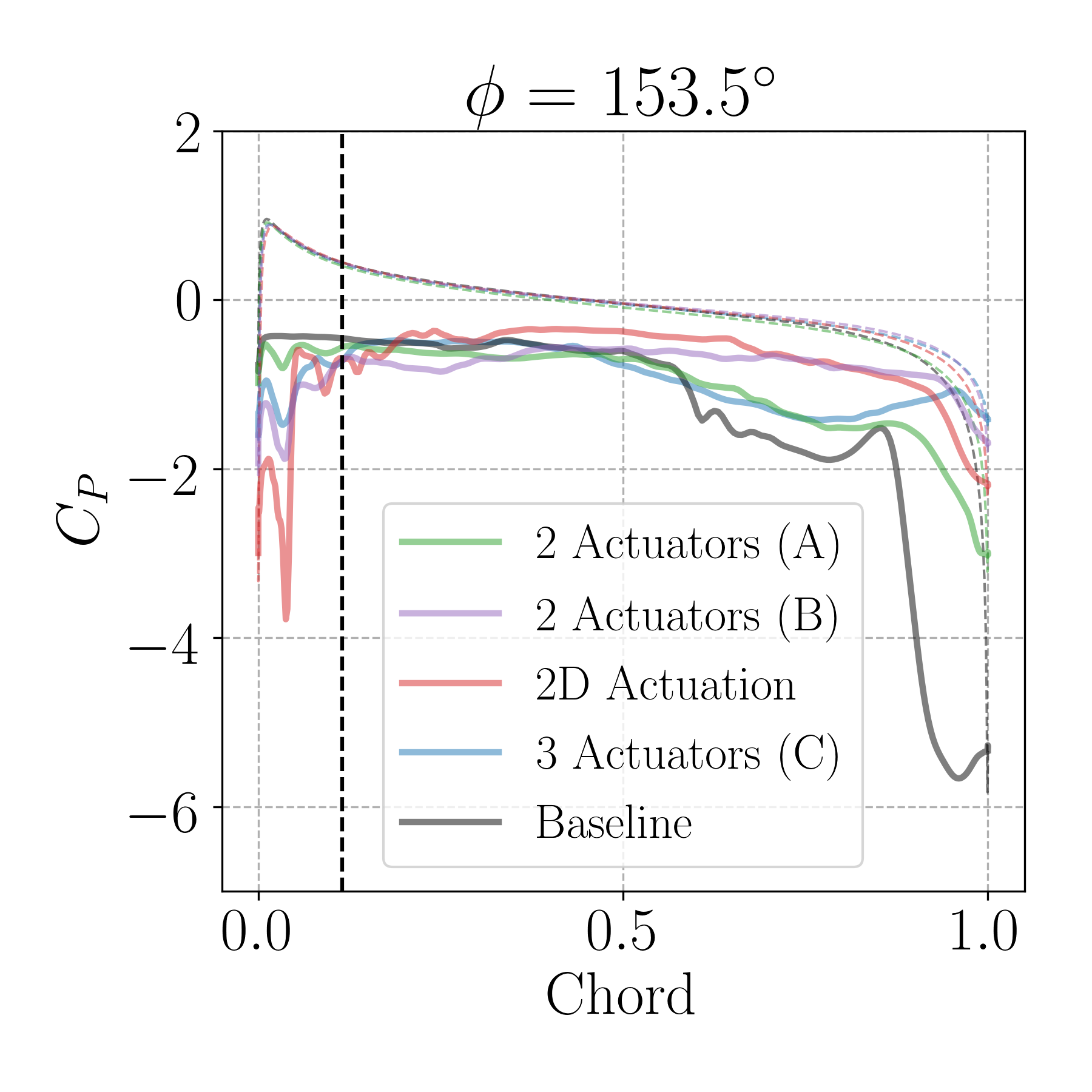}
		\end{subfigure}		
		\caption{Spanwise-averaged values of $ C_P $ for different configurations of actuation with $ St = 5  $ (Case 2). The vertical dashed line indicates the location of $ x_{vsn} $.}
		\label{fig:cpCurves3DActuators}
	\end{figure}

	Finally, Fig. \ref{fig:cpCurves3DActuators} presents a comparison of spanwise-averaged values of $ C_P $ for different configurations of actuation. One can see that the 2D actuation leads to lower values of $C_P$ at the leading edge, increasing lift and reducing drag. At the same time, the suction effects towards the trailing edge are milder for this case, further reducing drag. Values of mean lift to mean drag ratio, as well as aerodynamic damping, are displayed in Table \ref{tab:lift2drag} for different actuation setups. The best results of $\frac{\overline{C_L}} {\overline{C_D}}$ are found for the 2D actuation followed by that with two wider slots (B). The same trend is observed when analyzing values of aerodynamic damping. In summary, it can be observed that when a larger region on the leading edge is covered by the slots, making it more similar to a 2D configuration, the better the results are in terms of drag reduction and aerodynamic damping increase.
	\begin{table}[ht]
		\caption{Mean lift to mean drag ratios and aerodynamic damping for different actuation configurations with $ St = 5 $ (Case 2).}
		\label{tab:lift2drag}
		\vspace{-13px}
		\begin{tabular}{ccccc}
			\toprule \toprule
			& \textbf{2D Act.} & \textbf{2 Slots (A)} & \textbf{2 Slots (B)} & \textbf{3 slots (C)}  \\ \midrule
			$\frac{\overline{C_L}} {\overline{C_D}}$ & 20.48            & 12.03                & 19.13                & 14.73    \\
			$\Xi$ & 0.0122            & -0.0318              & 0.0066               & -0.0146   \\ \bottomrule \bottomrule
		\end{tabular}
		
	\end{table}
	\section{Conclusions}
	
	Large eddy simulations are conducted to study the flow over a SD7003 airfoil in a plunging motion. Results from the simulations are compared to data available in the literature for similar conditions and exhibit good agreement. In the current flow, instabilities arise after the beginning of the downstroke motion and a leading-edge vortex (LEV) is formed. Vorticity accumulates in the LEV, which reaches a given size, and is advected along the suction side of the airfoil increasing both lift and drag while reducing the pitching moment that induces a nose-down motion. Close to the trailing edge, the LEV is ``lifted" away from the airfoil surface by a trailing-edge vortex (TEV) that forms and is also advected. As the airfoil moves upward, the flow relaminarizes. Inherent variations from cycle to cycle occur due to turbulence that develops on the airfoil suction side and, thus, four from a total of five simulated cycles are phase-averaged to calculate aerodynamic loads. In general, good agreement is found between the phase-averaged quantities and those obtained from individual cycles.
	
	Simulations with 2D and 3D blowing and suction actuation are conducted for different frequencies which are characterized by Strouhal numbers $ St = 0.5 $ to $25 $. We also perform an assessment of flow actuation in terms of coefficient of momentum $ C_{\mu} $. Results demonstrate that actuation around $ St = 5 $ is effective in reducing both drag ($ C_D $) and quarter-chord pitching moment coefficients ($ C_M $) with only a mild loss in lift. For this specific frequency, it is shown that the dynamic stall vortex is broken into smaller coherent structures, leading to a pressure increase along the airfoil suction side, towards the trailing edge region. At the same time, pressure values on the suction side near the leading edge are considerably reduced, leading to a less severe lift loss and a further reduction in drag. Therefore, significant reduction in $ C_D $ and $ C_M $ are achieved as a result of mitigation of the dynamic stall vortex.
	
	Flow configurations with 3D actuation showed that, despite being able to mitigate some of the dynamic stall vortex effects, they are not as efficient in providing a high mean lift to mean drag ratio when compared to 2D actuation. In the 3D actuated cases, transition to turbulence occurs earlier compared to 2D actuation. This effect is due to formation of three-dimensional structures which do not severely impact the disruption of the LEV, differently than the 2D actuated flow. Nevertheless, all types of 3D actuation are able to modify the LEV sufficiently such that the TEV forms farther away from the trailing edge, diminishing its impact in the overall aerodynamic loads.The present study reveals that higher mean lift to mean drag ratios and aerodynamic damping are achieved when the actuator covers the whole airfoil span (2D actuation). Even when considering only actuators with variable spanwise widths and distribution, the most effective ones are those that cover the largest spanwise surface.
	
	\section*{Acknowledgments}
	
	BLOR and WRW acknowledge the financial support received from Funda\c{c}\~{a}o de Amparo \`{a} Pesquisa do Estado de S\~{a}o Paulo, FAPESP, under Grants No. 2013/08293-7, 
	2013/07375-0, 2016/24504-6 and 2018/04210-3. BLOR and WRW also thank CENAPAD-SP (Project 551), SDUMONT-LNCC and CEPID-CeMEAI for providing the computational resources used in the present simulations. CAY and KT acknowledges the support from US Air Force Office of Scientific Research under Grant No. FA9550-18-1-0040.

\bibliography{bibtex}

\end{document}